%% file: fgcs.tex
\documentclass[5p,times]{elsarticle}

\usepackage{ecrc}

\volume{00}
\firstpage{1}

\journalname{Future Generation Computer Systems}

\runauth{Wang, et al.}

\jid{FGCS}

\jnltitlelogo{FGCS}

\usepackage{amssymb}
\usepackage{graphicx}
\usepackage{amsmath}
\usepackage{algorithmic}
\usepackage{url}

\usepackage[numbers]{natbib}
\usepackage{textcomp}
\usepackage{xcolor}
\usepackage{hyperref}
\usepackage{xspace}
\usepackage{wasysym}
\usepackage{arydshln}
\usepackage{subfig}
\usepackage{amsthm}
\usepackage{amsfonts}
\usepackage{stfloats}
\usepackage{nccmath}
\usepackage[ruled,norelsize]{algorithm2e}
\usepackage{picins}
\usepackage{color,soul}

\usepackage{pgfplots}
\pgfplotsset{compat=1.10}
\usepgfplotslibrary{statistics}
\usepackage{tikz}
\usetikzlibrary{shapes}
\usetikzlibrary{matrix}
\usetikzlibrary{patterns}
\usetikzlibrary{backgrounds}
\usetikzlibrary{scopes}
\usepackage{multirow}

\newcommand{\neck}[1]{{\vspace{3pt}\noindent{\textbf{#1}}}}

\makeatletter
\newcommand{\removelatexerror}{\let\@latex@error\@gobble}
\makeatother

\usepackage{lineno}

\begin{document}

\begin{frontmatter}

\title{An Edge-Cloud Integrated Framework\\for Flexible and Dynamic Stream Analytics}

\author[a1,b1]{Xin Wang}
\address[a1]{Department of Information Systems, University of Maryland, Baltimore County, Maryland, United States}
\address[b1]{Center for Real-time Distributed Sensing and Autonomy, University of Maryland, Baltimore County, MD, United States}
\ead{xinwang11@umbc.edu}

\author[a1,b1]{Azim Khan}
\ead{azimkhan22@umbc.edu}

\author[a1,b1]{Jianwu Wang\corref{d1}}
\cortext[d1]{Corresponding author}
\ead{jianwu@umbc.edu}

\author[a1,b1]{Aryya Gangopadhyay}
\ead{gangopad@umbc.edu}

\author[c1]{Carl Busart}
\address[c1]{DEVCOM Army Research Laboratory, Adelphi, Maryland, United States}
\ead{carl.e.busart.civ@army.mil}

\author[c1]{Jade Freeman}
\ead{jade.l.freeman2.civ@army.mil}

\begin{abstract}
With the popularity of Internet of Things (IoT), edge computing and cloud computing, more and more stream analytics applications are being developed including real-time trend prediction and object detection on top of IoT sensing data. One popular type of stream analytics is the recurrent neural network (RNN) deep learning model based time series or sequence data prediction and forecasting. Different from traditional analytics that assumes data are available ahead of time and will not change, stream analytics deals with data that are being generated continuously and data trend/distribution could change (a.k.a. concept drift), which will cause prediction/forecasting accuracy to drop over time. One other challenge is to find the best resource provisioning for stream analytics to achieve good overall latency.
In this paper, we study how to best leverage edge and cloud resources to achieve better accuracy and latency for stream analytics using a type of RNN model called long short-term memory (LSTM). We propose a novel edge-cloud integrated framework for hybrid stream analytics that supports low latency inference on the edge and high capacity training on the cloud. To achieve flexible deployment, we study different approaches of deploying our hybrid learning framework including edge-centric, cloud-centric and edge-cloud integrated. Further, our hybrid learning framework can dynamically combine inference results from an LSTM model pre-trained based on historical data and another LSTM model re-trained periodically based on the most recent data. Using real-world and simulated stream datasets, our experiments show the proposed edge-cloud deployment is the best among all three deployment types in terms of latency. For accuracy, the experiments show our dynamic learning approach performs the best among all learning approaches for all three concept drift scenarios. 
\end{abstract}


\end{frontmatter}

\section{Introduction}

Stream analytics has become a major data analytics area due to hardware and software advances in Internet of Things (IoT), edge computing and cloud computing. It is now much easier to obtain sensing data from IoT devices, which leads to more and more stream analytics applications including real-time trend prediction~\cite{pandya2020adaptive,uchiteleva2021trils,qaisar2017fog} and real-time object detection~\cite{li2016deepcham,deepstream} on top of IoT sensing data. Different from traditional analytics that assumes data to be processed are available ahead of time and will not change, stream analytics processes data that are being generated on the fly and continuously. A well-known challenge of stream analytics is concept drift~\cite{vzliobaite2016overview,gama2014survey} which describes changes in the concept or distribution of stream data. 

There is a growing number of studies on how to conduct stream analytics by leveraging IoT, edge and cloud resources. Edge computing in an IoT environment brings computation and data storage closer to data sources. It operates on ``instant data" that is usually time sensitive. Besides the latency benefit, edge computing is normally designed for remote locations, where there is limited or no connectivity to a centralized computation location. However, resources on edges are constrained and limited in their capacity/capability and can only support relatively simple data processing like inference/prediction based on a pre-trained model. So it often relies on additional resources, such as storage or memory optimized devices, for more complex processing. Cloud computing~\cite{aws} provides on-demand computational resources for data analytics over the Internet and becomes a major approach for supporting complex and high-performance computation. Besides IoT environments, streaming data could also be delivered directly to the cloud and be computed with enough computing power and storage capacity. However, considering its distance to the data source, it is hard to have a quick response when data injection for some time-sensitive applications like earthquake warnings and automatic driving. Since both edge and cloud resources have their advantages and disadvantages, a related computing paradigm like Edge-to-Cloud Continuum~\cite{luckow2021pilot} has been proposed to integrate edge with cloud. In an edge-cloud integrated framework, the computation involves both front-end on-premise edge resources like Raspberry Pi and NVIDIA Jetson Nano, and back-end computing resources like big data and GPU clusters in cloud.

Deep learning has been widely used in stream analytics in IoT, edge or cloud environments. As a recent survey paper~\cite{mohammadi2018deep} shows, about one third of studies surveyed in the paper employs recurrent neural network (RNN) based deep learning models for time series or sequence data prediction and forecasting. RNN models can help learn temporal dependence and structures like trends and seasonality. Most existing studies and systems, such as~\cite{mohammadi2018deep} and~\cite{deepstream}, only support deep learning based inference on IoT/edge devices. A new research area is how to best integrate both edge resources and cloud resources for deep learning applications. Several researchers~\cite{luckow2021pilot,osia2018private,9615824,balouek2019towards,nezami2021decentralized,naha2020deadline,abdelbaky2017computing,murwantara2021adaptive} have proposed solutions and frameworks for streaming data analytics that leverage the capabilities of cloud services. However, to integrate edge with cloud, we need to achieve a proper trade-off between latency and accuracy for stream analytics between edge and cloud resources.

Accuracy and latency are two common metrics in stream analytics and many studies have how to balance them or make trade-offs. In the paper, we focus on how to achieve good accuracy and latency for the RNN-based deep learning model in an edge-cloud integrated environment by addressing the following two challenges. First, while the existing studies like~\cite{luckow2021pilot,osia2018private,9615824} provided promising direction, it is still not clear how to best deploy RNN-based deep learning models in edge and cloud resources for stream analytics to achieve better latency. Second, even though there have been many studies~\cite{zhang2017hybrid, pandya2020adaptive} on how to deal with unknown or changing data distributions in stream data, a.k.a. concept drift, it is still an open question how to balance accuracy and latency for RNN based stream analytics in an edge-cloud environment. 

To tackle the above two challenges, we propose a novel edge-cloud integrated framework and its corresponding open-source modules~\cite{Hybrid-Edge-Cloud} for stream analytics. To the best of our knowledge, our work is the first to achieve hybrid RNN-based deep learning for stream data in an edge-cloud integrated environment. Our contributions are summarized as follows.

\begin{itemize}
    \item We propose a novel edge-cloud integrated framework for stream analytics that supports low latency inference on the edge and high capacity training on the cloud. Tasks like data injection, model inference and synchronization are encapsulated as modules and can be flexibly deployed on either an edge device like Raspberry Pi or a cloud resource like AWS.
    \item Based on users’ preferences, we propose three flexible deployment modalities for our hybrid learning framework: edge-centric, cloud-centric and edge-cloud integrated. Based on a modular design, the hybrid learning framework can still work even if parts of the cloud services or edge analytics are unavailable. We further measured the latency differences between the three deployments using a real-world stream analytics application. Our experiments show the proposed edge-cloud deployment is among the best in terms of latency for inference, also will not run into capacity limitation for training.
    \item To adapt the concept drift challenge of stream data in edge-cloud integrated environments, we propose an adaptive hybrid learning framework that combines and benefits from both cloud resources’ high capacity and edge resources’ low latency. Our hybrid learning framework contains batch learning by employing a pre-trained RNN model from large historical data, speed learning by periodically re-training an RNN model from most recent data and hybrid learning by combining predictions from batch and speed learning. We also study a new hybrid learning algorithm that can combine results dynamically. Our experiments show our hybrid learning approaches can have better RMSE than cloud-based batch learning and edge-based speed learning in most cases and our dynamic learning approach performs the best among all learning approaches for all three concept drift scenarios.  
\end{itemize}

The rest of the paper is organized as follows. In Section \ref{sec:background}, we briefly introduce the related background our work is built on. Section \ref{sec:overview} provides an overview of the proposed edge-cloud integrated hybrid learning framework. Section \ref{sec:flexible} introduces our three flexible deployment modalities of hybrid learning framework, including edge-centric, cloud-centric and edge-cloud integrated deployments. The adaptive hybrid stream analytics and its two weight combinations, namely static and dynamic weighting algorithms, are explained in Section \ref{sec:adaptive}. Evaluations and benchmarking results are next discussed in Section \ref{sec:eva}. We summarize related studies and compare them with our work in \ref{sec:relatedworks} and conclude in Section \ref{sec:conclusions}.

\section{Background} \label{sec:background}
\subsection{Edge Computing and Edge-cloud Integration}
Applications that utilize IoT devices are increasing day by day and data volumes produced by IoT edge devices could be enormous. In order to alleviate the heavy load of data transfer, edge devices can pre-process, analyze and quickly react to the time-sensitive application near data sources, and only deliver the processed data or inference results to back-end computation centers. So, when data is handled by an edge device that is close to data generation source, we could achieve faster response time, higher computing efficiency and lower network traffic in comparison to the case where IoT data is processed in a centralized computation location. However, the capacity of edge devices limits their capability to handle complex heterogeneous data and even could lead to unacceptable and unpredictable performance. To deal with the challenge, work at~\cite{balouek2019towards} extended the idea of computing continuum, and proposed an edge-to-cloud integration to support dynamic and data-driven application workflows, which are capable of reacting to unpredictable and heterogeneous real-time data. Pilot-Edge~\cite{luckow2021pilot} proposed its abstraction to support data and machine learning (ML) applications in the edge-to-cloud continuum, which was designed to address the challenge of computation performance in heterogeneous edge environments.

\subsection{RNN Models on Edge Devices}
In recent years deep learning (DL) has gained attention due to its ability to facilitate analytics in the IoT domain~{\cite{mohammadi2018deep}}. Sequence DL models like RNN and LSTM are useful for streaming data prediction since these models can learn hidden features from a sequence of records. Hermans et al.~{\cite{hermans2013training}} state that considering the architecture and the functionality of RNNs, the hidden layers in RNNs are supposed to provide a memory instead of hierarchical processing of features. LSTM, as a special form of RNN, uses the concept of gates to actively control the memory cell and prevent perturbation from irrelevant inputs. The work by Chung et al.~{\cite{69e088c8129341ac89810907fe6b1bfe}} show that LSTM models perform better than RNN models when data is characterized by a long dependency like the observations from IoT applications. Tao et al.~{\cite{tao2016multicolumn}} use LSTM architecture and mobile phone sensor data for human activity recognition. 

More advanced sequence models have also been proposed for stream analytics. Zhang et al.~{\cite{zhang2019novel}} propose a multi-head convolutional neural network with multi-path attention to detect human activity signals received from the wearable sensors. These experiments are carried out on a local computer rather than the edge device and the authors mention their attention models are computationally expensive.

The above studies only support a pre-trained RNN model. As an initial work that supports RNN model update based on more recent data, in this paper, we study a lighter weight LSTM model on the edge device. In future works, we will explore more complicated sequence models which use attention and study how to best enable model inference and updates with limited resources at the edge.

\subsection{Concept Drifts in Real-world IoT Data Streams}
In real-world data-driven applications, analytics of IoT streaming data often encounters the change in the data distribution while extracting different features from stream sources. These hidden changes in the concept or distribution of streaming data, which are unknown to the learning algorithms, are termed as concept drift~\cite{vzliobaite2016overview,gama2014survey} or nonstationary data. Mathematically, if we denote $X$ as an input vector and $y$ as an output vector, then $(X,y)$ will be an infinite sequence of data streams. Concept drifts between time point $t_i$ and time point $t_j$ can be defined as

\begin{equation} \label{equ:drift}
p_{ti}(X,y) \neq p_{tj}(X,y)
\end{equation}

where $p_{ti}$ and $p_{tj}$ denote joint probability distribution at time $t_i$ and $t_j$, respectively.

Changes in streaming data distribution over time might appear in various ways such as gradual drift and abrupt drift. Abrupt drift happens suddenly by switching from one concept to another in any time period~\cite{gama2014survey}. Gradual or incremental drift does not change abruptly, instead happens over a long period and therefore can be expected. It defines a continuous change that happens from one underlying process behavior to another one. In this paper, simulated datasets consisting of gradual and abrupt drift were used to know how hybrid stream analytics reacts in the context of different types of drifts.

\subsection{Adaptive Learning and Lambda Architecture}
Because underlying concepts of real-world stream data could evolve over time, adaptive learning algorithms have been proposed to address concept drift by adapting new instances and forgetting old ones in order to naturally follow drifts in the stream. It can also be considered as improved incremental learning algorithms that are able to integrate fresh data during their operation to react to concept drifts~\cite{gama2014survey}. Mentioned by~\cite{krawczyk2018online}, concept drift detector, sliding windows, online learner and ensemble learners are the most common adaptive learning approaches. One challenge is, the estimation of performance feedback is difficult for any adaptive learning system due to the absence of ground truth in stream data. Besides, the anomalies of the algorithm can readily be confused for changes in the stream data. In our paper, we infer and evaluate our adaptive learning method by analyzing earlier historical data and the data in the past time windows.

Lambda architecture is a data-processing design pattern which is usually used in data-driven applications by taking advantage of both batch and stream processing methods~\cite{pandya2020adaptive}. The lambda architecture has three layers, batch layer for batch processing based on historical data, speed layer for real-time stream processing, and serving/hybrid layer for combining outputs from both batch and speed layer. The goal for lambda architecture design pattern is to abstract and balance both the accuracy by using batch processing to provide comprehensive knowledge from historical data, and the latency by using stream processing to learn the resent changes from real-time data. Inspired by this design pattern, we propose a hybrid stream learning model to achieve adaptive learning.

\section{Overview of Hybrid Stream Analytics Framework} \label{sec:overview}
In this section, we briefly introduce our proposed hybrid stream analytics framework from a high-level view. By combining edge resources' lower communication latency with cloud resources' higher computational power, we propose a novel hybrid learning framework that can achieve good latency and accuracy for stream analytics. 

\begin{figure}[h]
\begin{center}
    \includegraphics[width=0.48\textwidth]{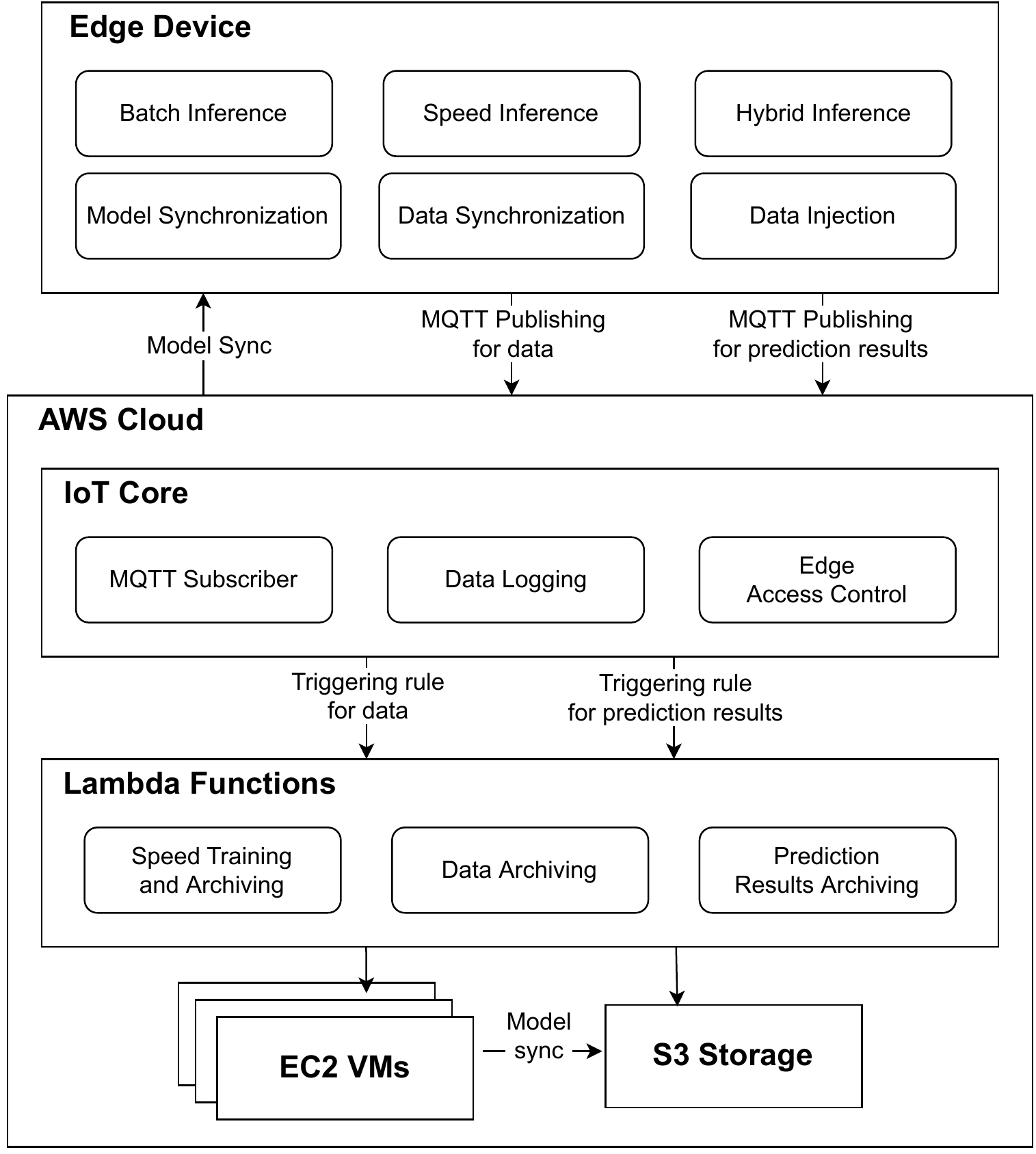}
    \caption{The overview of our proposed hybrid stream analytics framework.}
    \label{fig:arch}
\end{center}
\end{figure}

We summarize our hybrid stream analytics framework in Figure \ref{fig:arch}. In our design, different functionalities are wrapped into multiple modules, which can be deployed on either edge or cloud. Within the edge side, six modules are designed for flexible and dynamic stream analytics. The first three modules, namely \textit{batch inference}, \textit{speed inference} and \textit{hybrid inference}, are the main functionality for the inference task of stream analytics. When stream data is injected, batch inference provides batch predictions based on a pre-trained model from historical data; speed inference enables predictions based on the latest model trained from the previous time window; and hybrid inference combines their inferred values to get a new prediction value. We will explain in detail how we leverage our hybrid learning model to achieve adaptive prediction further in Section \ref{sec:adaptive}. Next, we introduce the last three modules on the edge side. For \textit{model synchronization}, it synchronizes the models for speed inference from cloud to edge periodically. For \textit{data synchronization}, it synchronizes the streaming raw data and all inference results to the cloud storage. All the synchronizations are achieved through the edge-cloud MQTT messaging~\cite{mqtt} based on specific topics. Module \textit{data injection} acts as a transfer station to throttle the amount of streaming data in each time window and control them to the target modules. All these modularized functionalities can work both independently and cooperatively based on usage. With this modular design, our hybrid stream analytics framework can still work even if some modules are unavailable. Details of how our framework achieves flexibility with different deployment modalities including edge-centric, cloud-centric and edge-cloud integrated scenarios will be explained in Section~\ref{sec:flexible}.

\begin{figure*}[t]
\begin{center}
    \includegraphics[width=0.55\textwidth]{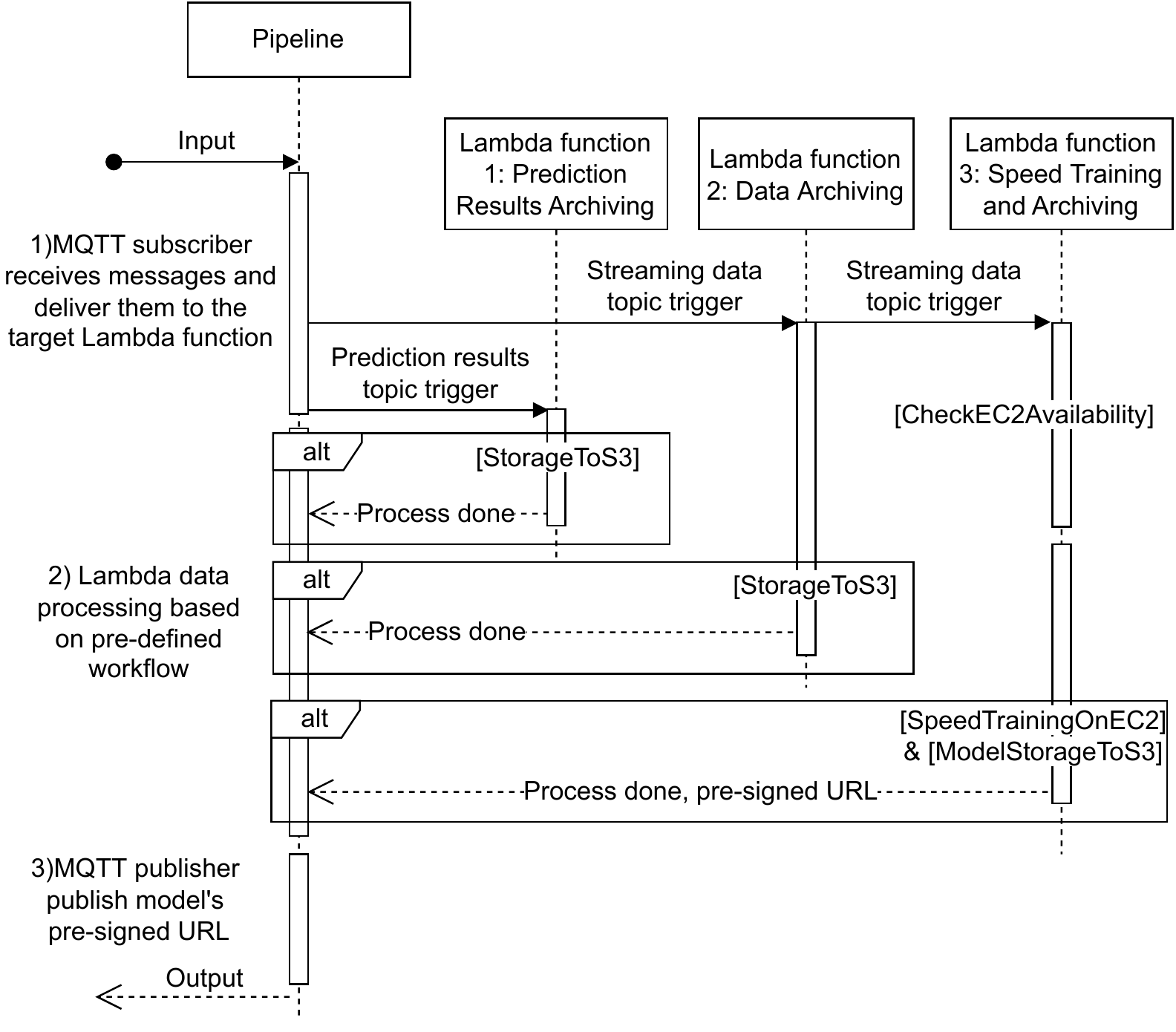}
    \caption{The system sequence diagram for pipelines of Lambda functions for back-end processing in the cloud.}
    \label{fig:lambda}
\end{center}
\end{figure*}

Within the cloud side, two resources are used as the back-end of the hybrid stream analytics framework: 1) AWS IoT Core manages the edge-cloud communication and accession, and 2) AWS Lambda Function implements the pipeline of complex data processing. AWS IoT Core provides resources and services that help users achieve edge-cloud computing with AWS IoT-based solutions. Within AWS open source IoT edge runtime, called Greengrass \cite{greengrass}, our pre-built modules can be deployed, communicated and managed on the edge through AWS web console or command line. Specifically, defined by AWS access control, all permitted edge-to-edge and edge-to-cloud communications are achieved by MQTT publishing and subscribing protocol. Besides, IoT Core enables triggering rules that provide a SQL-based language to filter MQTT payloads and deliver them to the target services like Lambda Function. As shown in Figure~\ref{fig:lambda}, filtering by the Lambda triggering rules, the incoming MQTT payloads will be delivered to different target Lambda functions as Lambda events (Step 1). In Step 2, Lambda functions will execute these triggered events asynchronously based on their pre-defined pipeline. \textit{Prediction Results Archiving} function only receives events from the inference results topic and directly stores the payloads to AWS S3 object storage. \textit{Data Archiving} and \textit{Speed Training and Archiving} functions both receive events from the streaming data topic. For \textit{Data Archiving} function, just like the first function, it stores the payloads to S3 directly. For \textit{Speed Training and Archiving} function, it will first check AWS EC2's availability, deliver streaming data to an EC2 virtual machine for model training, and then upload the latest model to S3 when training finished. In the meanwhile, in Step 3, this Lambda function will also publish a one-time pre-signed S3 URL to the edge. This S3 URL is signed with cloud credentials, which grants temporary access to the edge's model synchronization.

\section{Flexible Deployments of Hybrid Stream Analytics Framework}\label{sec:flexible}
\begin{figure*}[ht]
\begin{center}
    \includegraphics[width=0.71\textwidth]{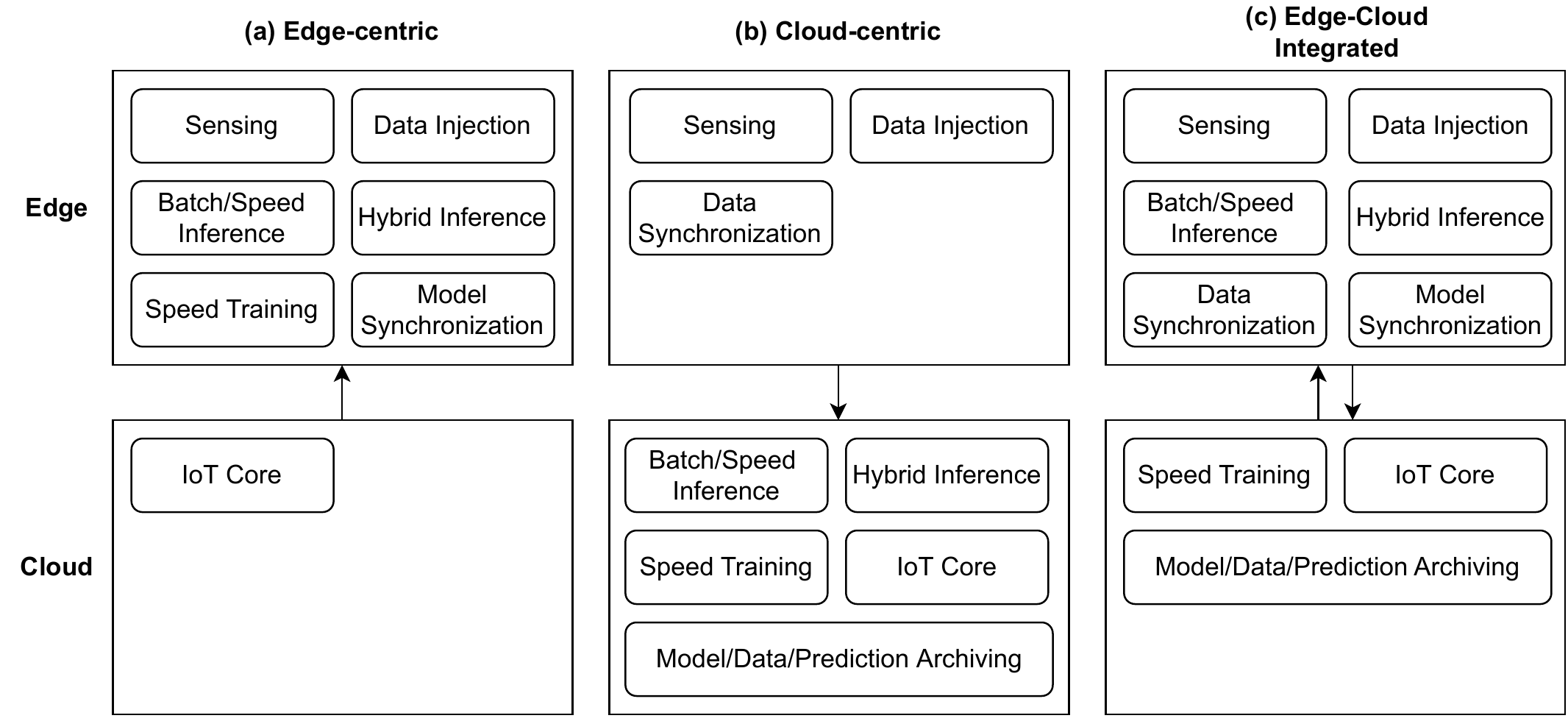}
    \caption{Flexible deployments of our hybrid stream analytics framework.}
    \label{fig:flexible}
\end{center}
\end{figure*}

To achieve the flexibility of the proposed hybrid stream analytics framework, we use a modular design for all framework components, which achieves a proper trade-off between latency and accuracy for stream analytics.
Based on different scenarios, we design three types of deployments for the hybrid stream analytics: edge-centric, cloud-centric and edge-cloud integrated deployments. The summary of the three deployment modalities is shown in Figure \ref{fig:flexible}. We also summarize the advantages and limitations of the proposed deployments in Table \ref{tab:flexible}.

\begin{table*}[hb]
\centering
\footnotesize
\caption{Advantages and limitations of the proposed three deployment types for stream analytics.}
\setlength{\tabcolsep}{0.99mm}{
\begin{tabular}{|c|c|c|}
\hline
\textbf{} & \textbf{Advantages} & \textbf{Limitations} \\
\hline
\hline
\multirow{2}{*}{Edge Centric} & quick respond since & ~capacity and capability shortage~ \\
& ~the computation is near to the source of data~ & for edge device \\
\hline
\multirow{2}{*}{Cloud Centric} & enough capacity and capability & high communication overhead \\
& for high accuracy computation & between edge and cloud \\
\hline
\multirow{2}{*}{~Edge-cloud Integrated~} & quick respond for inference and & complex coordination \\
& high accuracy for training & between edge and cloud \\
\hline
\end{tabular}}
\label{tab:flexible}
\end{table*}

\subsection{Edge-Centric Stream Analytics Deployment}
Because stream analytics needs to process incoming data continuously, it is common that the back-end cloud service will be temporally unavailable due to network disconnection or resource overload problems. For this, we design an edge centric deployment modality, which allows the edge to execute stream analytics autonomously with local events, as shown in Figure \ref{fig:flexible}a. 

We can summarize the unavailability of the cloud into two scenarios: part of cloud computational resources (like EC2) is unavailable and the whole cloud service is unavailable. For the first scenario, IoT Core and other Lambda functions still work well except for the \textit{Speed Training and Archiving}. 
If the Lambda function cannot connect to any virtual machine in EC2, it will put the process event into its waiting queue and wait for the available resources. At this time, although other services still work fine, the performance of speed inference may not have good accuracy since it still uses an ``out-of-date" model trained by the data from the time window before the unavailability of EC2.

When the whole cloud service is unavailable, that means the edge cannot get any update from the cloud. In this scenario, all the flexible modules and the functionalities of Lambda Function are wrapped into AWS Greengrass runtime and deployed to the edge side. Based on the usage, if the stream analytics framework is only assigned to do model inference, the batch, speed and hybrid inference modules will wait for the data from data injection and output the results separately. If the stream analytics framework also requires to do model training, the speed training module can be deployed to the edge device, subscribe to the data injection, fulfill speed training and synchronize the new model for next time window inference. To achieve this usage, MQTT publishes and subscribes messages locally within the edge device. Specifically, the speed training module is in a containerized Spark~\cite{spark-url} based design. By default, it will initiate the training environment from a pulled docker image and allocate available edge resources to train the model in a Spark standalone mode. Our future work will study how this Spark-based speed training module can be extended to different edge devices as a distributed master-worker computing. If there are idle edge devices, Spark will initiate these edges as workers and auto-scale the training tasks to them. This Spark-based parallel design of the speed training module also avoids the issue of limited computational capability of individual edge devices.

\subsection{Cloud-Centric Stream Analytics Deployment}\label{sec:flexible_cloud}
To achieve the flexibility of the hybrid stream analytics framework, we also provide a cloud-centric deployment for stream analytics. In this scenario, as shown in Figure \ref{fig:flexible}b, the edge device is only used to sense the streaming data and synchronize them to the cloud. 

When the edge device cannot perform any data processing, the batch, speed, hybrid inference and model synchronization modules should be deployed on cloud computational resources like EC2. At this time, IoT Core service will mark the EC2 virtual machine (VM) as the substituted edge computational capability and subscribe to the MQTT payloads from EC2. 

Leveraged by Lambda functions, our cloud-centric deployment achieves automatic data processing for each time window of stream analytics in the back-end cloud. Like the introduction in Section \ref{sec:overview}, for the hybrid stream analytics framework, when an incoming MQTT payload triggers a filtering rule, IoT Core invokes corresponding Lambda functions asynchronously and passes the data payload from edge to the specific function. After that, based on the pre-defined functions of data processing pipeline, Lambda will check the availability of AWS EC2 virtual machines, train the streaming data in EC2 and archive the model to the S3 object storage. Later, Lambda will reply with a one-time pre-signed S3 URL to edge, which grants the model synchronization module temporary access to synchronize the latest model from cloud.

\begin{table*}[b]
\centering
\caption{Cloud service mapping for hybrid stream analytics. }
\resizebox{.99\textwidth}{!}{%
\begin{tabular}{|c|c|c|c|c|}
\hline
\textbf{Service category} & \textbf{Service description} & \textbf{Amazon AWS} & \textbf{Microsoft Azure} & \textbf{Google Cloud}\\ \hline \hline
 Virtual machine & Virtual instance that enables to host speed training. & EC2 & Virtual Machines & Compute Engine\\ \hline
 IoT platform & Manage the edge-cloud communication and accession. & IoT Core & IoT Hub & Cloud IoT\\ \hline
 IoT runtime & Help edges to build, deploy and manage the application. & Greengrass & IoT edge & Cloud edge\\ \hline
 Container service & Store, manage, and secure container images in private or public. & ECR & Azure Container Registry & Artifact Registry\\ \hline
 Object storage & Store, manage, and secure any amount of data in storage. & S3 & Blob storage & Firebase\\ \hline
 Serverless & Run and manage the application with zero server management. & Lambda Functions & Azure Functions & Cloud Functions\\ \hline
 Cloud Python SDK & Easy-to-use interface to access cloud services. & Boto/Boto3 & .NET Core & Cloud SDK \\ \hline
\end{tabular}
}
\label{tab:extensibility}
\end{table*}

\subsection{Edge-Cloud Integrated Stream Analytics Deployment}
We first make a short summary about the advantages and limitations of the first two proposed deployments, as shown in Table \ref{tab:flexible}. For the edge-centric deployment, although all the hybrid stream analytics can run closer to the data sources, the limitation is that the weak computational capability and capacity of edge devices may cause process congestion or even crash during stream analytics. Whereas for the cloud-centric deployment, all the data cannot be pre-processed before it arrivals to cloud. In another word, the edge-centric deployment focus on the quick on-site response of the stream analytics and the cloud-centric deployment mainly benefits of the computing power and storage capacity of cloud. 

In order to achieve the proper trade-off between these two deployment modalities, we propose a third deployment modality, namely edge-cloud integrated deployment. As shown in Figure \ref{fig:flexible}c, with edge-cloud integrated deployment, all the inference and synchronization modules are developed on the edge, while speed training and all the archiving are developed on the cloud. With this edge-centric deployment solution, hybrid stream analytics can enjoy not only the computing power and storage capacity of cloud, but also the low latency for edge resources.

\subsection{Flexible Deployment of Our Framework}
In our hybrid stream analytics framework, we have six modules implemented as Python functions. For the flexible deployment, we use different ways to wrap these modules to achieve proper coordination and trigger their invocations based on incoming stream data. Specifically, we use AWS IoT Component with its Greengrass runtime for edge-based deployment and AWS Lambda function for cloud-based deployment. To deploy a module on edge, AWS IoT Component is required with an update interval configuration so the modules can be triggered by an AWS IoT event and the records of the time windows periodically. To deploy a module on AWS, it can be encapsulated into the docker container with its software environment. In this way, the same modules and implementations can be reused when switching from one deployment to another.

\subsection{Extensibility of Our Framework}
In this paper we implement the hybrid stream analytics framework on AWS cloud, however the proposed framework can be easily extended to other cloud providers. Most services from different cloud providers can be mapped to each other. Table {\ref{tab:extensibility}} lists related cloud services provided by Amazon AWS, Microsoft Azure and Google Cloud for hybrid stream analytics. In order to achieve extension to Microsoft Azure and Google Cloud, the user needs to wrap the flexible modules into its corresponding IoT runtime for each cloud. All functionalities can be wrapped into the Greengrass runtime for AWS, the IoT Edge runtime for Azure, and IoT Cloud Edge for Google. Additionally, the serverless functions are needed to adapt to the specific structure and format of each cloud.

\section{Adaptive and Dynamic Hybrid Learning Model for Stream Analytics} \label{sec:adaptive}

In order to tackle the challenge of concept drift, we propose a hybrid learning model, which can adapt to the changes in stream analytics by weighted combining the results from batch and speed inference. Like the design pattern of Lambda architecture, hybrid stream analytics should contain a batch layer, a speed layer and a serving/hybrid layer. In this section, we will first provide an overview of our hybrid learning model. Then, we introduce the orchestration of the hybrid stream analytics, and its two weight combination algorithms, namely static weighting algorithm and dynamic weighting algorithm.

\subsection{Overview of Adaptive Hybrid Stream Analytics} \label{sec:statement}
Leveraging the lambda architecture, our hybrid stream analytics achieves adaptability of stream data concept drift. We first introduce problem statements of our hybrid stream analytics. In our hybrid learning model, we separate the inference tasks of stream analytics into three layers: batch layer, speed layer and hybrid layer.

\neck{Batch layer tasks.} For the batch layer, our hybrid learning model only trains the model once and reuses it for inference all received stream data. Its model is defined as

\begin{equation} \label{equ:statement_batch}
\hat y^{i} = f(y^{i-1}, y^{i-2}, \dots, y^{i-n})
\end{equation}

We call the training in batch layer as the batch training and its inference as batch inference.

\neck{Speed layer tasks.} For the speed layer, there is no pre-trained model before the stream analytics begins. Instead, the speed layer re-trains a new model for every time window and uses it to infer the next time window data. We define the inference task as follows. For each time window $t$, the stream analytics trains a model $f_t$ and uses it to make predictions for the new time window $t+1$. For each timestep $i$ within time window $t+1$, the prediction value $\hat y_{l}$ can be defined as 

\begin{equation} \label{equ:statement_speed}
\hat y^{i}_{t+1} = f_t(y^{i-1}_{t+1}, y^{i-2}_{t+1}, \dots, y^{i-n}_{t+1})
\end{equation}

We call the training in speed layer as the speed training and its inference as speed inference. 

\neck{Hybrid layer task.} For hybrid layer, in order to aggregate the inference results from both consistent patterns of historical data distribution and the hidden changes of streaming data distribution, its model works based on formula

\begin{equation} \label{equ:hybrid}
Pred_{hybrid}= W^s * Pred_{speed} +  W^b * Pred_{batch}
\end{equation}

where the weights $W^s+W^b=1$. The hybrid layer only has inference (no training), so that we call it as hybrid inference.

\begin{figure*}[ht]
\begin{center}
    \includegraphics[width=0.76\textwidth]{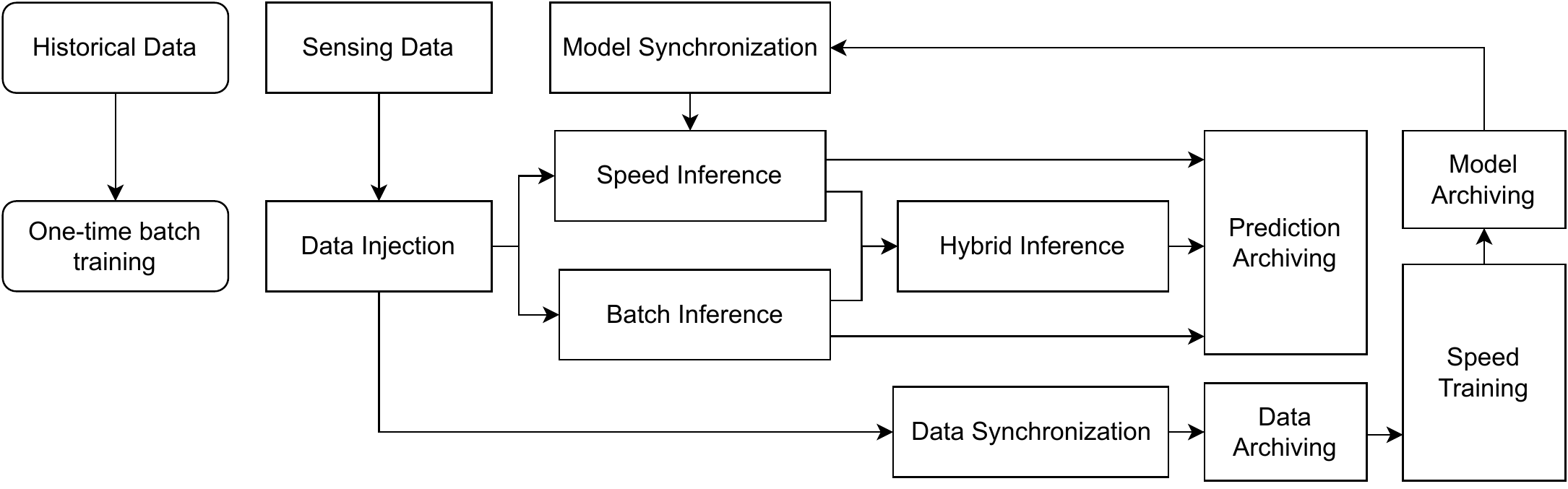}
    \caption{Module orchestration of our hybrid stream analytics (rectangular boxes denote periodic operations and round boxes denote one-time operations).}
    \label{fig:workflow}
\end{center}
\end{figure*}

Because model training happens only once for the batch layer, referred to as Figure \ref{fig:workflow}, the latency of batch training is not part of the latency occurred for incoming streaming data. Instead, we only focus on the latency for batch inference, speed training, speed inference and hybrid inference for each time window in the paper. Also, we run each module asynchronously to lower overall latency.

\subsection{Orchestration of Hybrid Stream Analytics}
We explain how the modules of the hybrid stream analytics orchestrate, which is illustrated in Figure \ref{fig:workflow}. There exist a one-time batch training before the stream analytics start.
After that, the data injection module acts as a transfer station to throttle the amount of sensing streaming data into a payload for each time window, for example, catching streaming data every 30 seconds. With data injection throttling, incoming streaming data can be temporarily stored in a buffer queue which avoids the receiver from the crash when absorbing the peaks of incoming data for a very short time lapse. Then, the data injection module delivers data based on the usages of stream analytics, which contains two asynchronous phases: training phase and inference phase.

In the training phase, stream analytics executes the speed training based on the stream data in each time window. After receiving raw streaming data, the speed training module trains a new model based on the current payload batch. Then this new trained model will be synchronized to the speed inference module for the prediction of the later stream analytics. 

In the inference phase, the stream analytics pipeline requires batch and speed inference, and the data injection module will deliver the streaming payload to the inference modules in each time window. With the design pattern of lambda architecture, batch inference module predicts results using a pre-trained model as the batch layer, which is learned from the historical dataset. As the speed layer, speed inference module updates its model for each time window, which uses a model learned from the previous time window and tests it in the current time window. And hybrid layer will aggregate the inference results both from the consistent patterns of historical data distribution and the hidden changes in streaming data distribution in both batch and speed layers. The hybrid inference results will later be published by the MQTT publisher for archiving and further notification.

\subsection{Hybrid Learning Model with Weight Combination} \label{sec:dynamicweights}
In this section, we introduce the two weight combination algorithms for hybrid inference.

\neck{Static Weighting Algorithm.}
The weight combination algorithm can be defined as the static weighting algorithm if the weights $W^s$ and $W^b$ (in Equation \ref{equ:hybrid}) are been set as a fixed value for every time window of the stream analytics. Because of the fixed weights, it is obvious that the hybrid learning model with the static weighting algorithm is hard to adapt to the dynamic changes of streaming data. To solve this problem, we provide an optimized approach to dynamically learn the weights during stream analytics, namely dynamic weighting algorithm. 

\neck{Dynamic Weighting Algorithm.} 
In theory, finding the dynamic weights is a mathematical optimization problem that is used to find the best solution from all feasible solutions. Shifting it to a machine learning problem, stacking ensemble methods~\cite{1688199,rokach2010ensemble} combine multiple machine learning algorithms to obtain a better predictive performance than that could be obtained from any of the constituent learning algorithms alone. Based on these ideas, we propose our dynamic weighting algorithm.

\begin{figure}[ht]
\removelatexerror
\vspace{-3pt}
\begin{algorithm}[H]
\small
\caption{Dynamic Weighting Algorithm (DWA)} \label{alg:dynamicweights}
\DontPrintSemicolon
\KwIn{$M^b$, $M_{t-1}^s$, $X_{t-1}^{test}$, $Y_{t-1}^{test}$}
\KwOut{$W^b_t$,$W^s_t$}

  \SetKwFunction{FMain}{\textbf{function} DWA}
  \SetKwProg{Fn}{}{:}{}
  \;
  \Fn{\FMain{}}{
  $EnsembleModels \gets [~]$\;
  $Pred \gets [~]$\;
  $EnsembleModels.append(M^b,M_{t-1}^s)$\;
  \;
  \For{$model$ \textbf{in} $EnsembleModels$}{
  $Pred.append(model.predict(X_{t-1}^{test}))$
  }\;
  $W^{init} = [0.5]*len(Pred)$\;
  $cons = lambda~W: 1-sum(W)$\;
  $bounds = [(0,1)]*len(Pred)$\;
  $loss = LossFunc(Y_{t-1}^{test},Pred)$\;
  $W^b_t,W^s_t \gets~minimize(loss, W^{init}, bounds, cons, Solver)$\;
  \;
  \KwRet $W^b_t,W^s_t$\;
  }
\end{algorithm}
\vspace{-3pt}
\end{figure}

As shown in Algorithm \ref{alg:dynamicweights}, for each time window $t$, taking the inputs of batch layer model $M^b$, speed layer model $M_{t-1}^s$ at time window $t-1$ with the test dataset $X_{t-1}^{test}$, the dynamic weighting algorithm stacks the provided models and collects their predictions using the test dataset as the serving layer. By listing the constraints and bounds, like limiting the sum of weights to equal 1 ($W^b_t+W^s_t=1$) and limiting the weights $W^b_t,W^s_t$ in range $[0,1]$, the optimization solver will find the optimum values that can minimize the objective loss function, starting from an initial guess $W^{init}$ (we choose 0.5 as our initial weights). In our paper, we use Sequential Least Squares Programming (SLSQP)~\cite{kraft1988software} as the optimization solver $Solver$, which is always used to solve nonlinear programming (NLP) problems. We also use Root Mean Squared Error~\cite{chai2014root} regression loss as our loss function $LossFunc$ that can be defined as 

\begin{equation} \label{equ:rmse}
L_{rmse}(y)=\sqrt{\frac{1}{n}\sum_{j=1}^{n}(y_j-\hat{y}_j)^2}
\end{equation}

which is the square root of the average of squared differences between prediction $\hat{y}_j$ and actual observation $y_j$.

In our algorithm implementation, at each time window $t$, we stack two pre-trained models in the serving layer, which include one speed layer model at time window $t-1$ and the batch layer model. Since the ensemble method does not require a constant pattern for stacking models from batch or speed layer, the dynamic weighting algorithm also has its variants like stacking the most resent $n$ speed layer models or stacking speed layer models continuously. We will study these variants as part of our future work.

\begin{figure*}[ht]
	\centering
	\subfloat[\textbf{No-Drift} Data (for all five variables). ]{
		\includegraphics[width=0.33\linewidth]{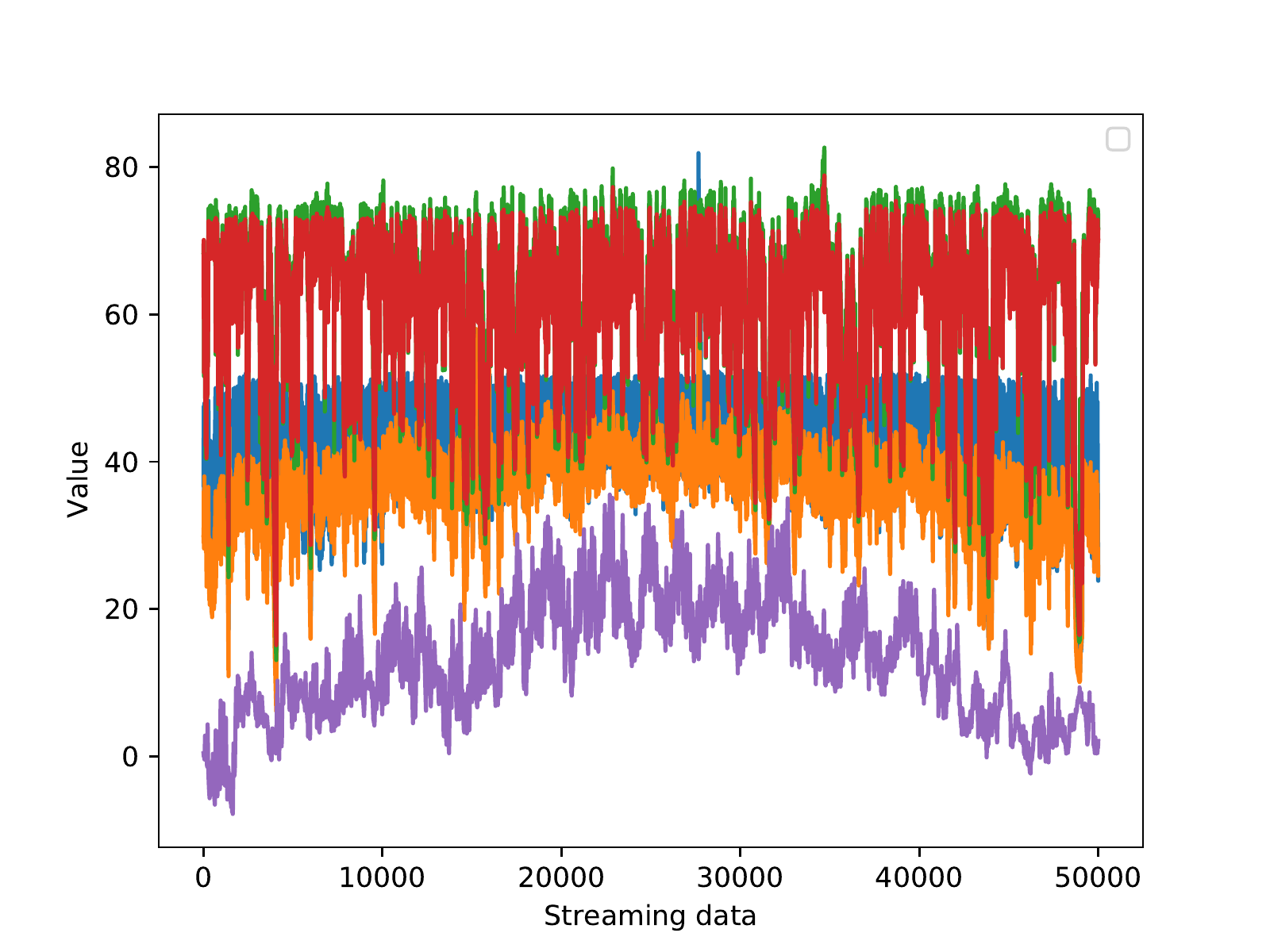}
		\label{fig:sourcedata}
	}
	\subfloat[\textbf{Gradual-Drift} Data (for target variable). ]{
		\includegraphics[width=0.33\linewidth]{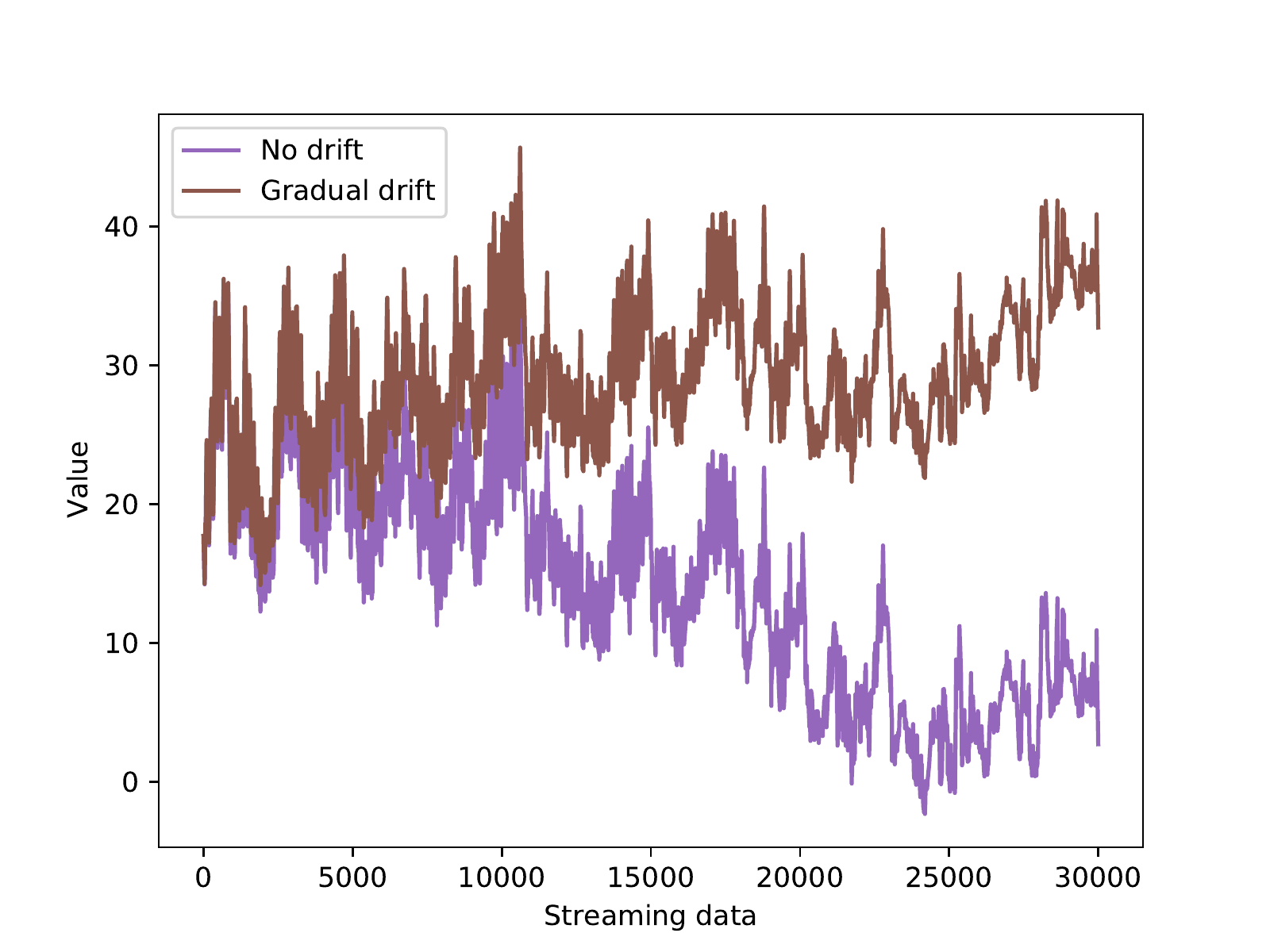}
		\label{fig:gradualdriftdata}
	}
	\subfloat[\textbf{Abrupt-Drift} Data (for target variable). ]{
		\includegraphics[width=0.33\linewidth]{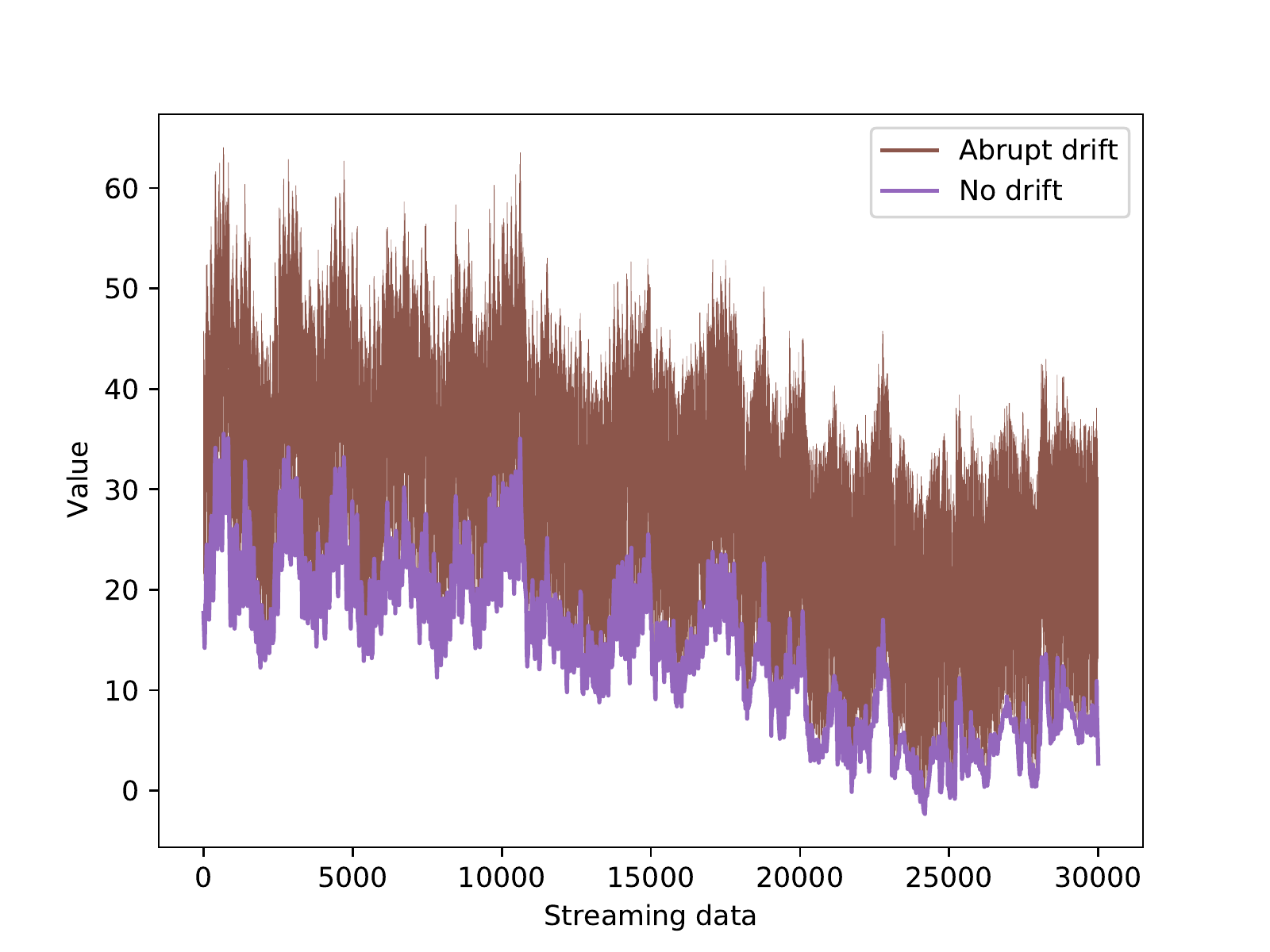}
		\label{fig:abruptdriftdata}
	}
	\caption{Data distribution for one real-world and two synthetic time-series of wind turbine temperature.}
	\vspace{12pt}
	\label{fig:data_distribution}
\end{figure*}

\section{Evaluation}\label{sec:eva}
This section conducts the evaluation of our proposed flexible and dynamic hybrid stream analytics framework. We implemented our framework and open-sourced it on GitHub at~\cite{Hybrid-Edge-Cloud}. One real-world and two synthetic datasets are applied in our experiments and the metric includes latency and accuracy. The evaluation compares the difference in two aspects: 1) different types of hybrid learning framework deployment and 2) different types of hybrid stream analytics approaches.

We note that our experiments have implemented approaches from several related works as baselines for specific capabilities: 1) For hybrid time series forecasting, we implemented the static approach in reference~{\cite{pandya2020adaptive}} as a baseline to evaluate the advantage of our proposed dynamic capability (see Section {\ref{sec:6.3.2}} and Figure {\ref{fig:rmse_boxplot}}); 2) For hybrid inference, we implemented reference~{\cite{uchiteleva2021trils}} as the baseline for batch inference modality to evaluate the advantages of our dynamic hybrid inference modality (see Section {\ref{sec:6.2}} and Table {\ref{tab:latency_inference}}).

\subsection{Datasets and Evaluation Settings}\label{sec:dataset}

\subsubsection{Dataset description} 
We use one real-world dataset and two simulated datasets for gradual drifts and abrupt drifts, in order to evaluate our proposed edge-cloud integrated framework. The data distributions of each dataset are shown in Figure \ref{fig:data_distribution}.

\neck{One real-world dataset.} Our application is designed for the real-world prediction of wind turbine temperature, based on the ENGIE’s open wind farm data~\cite{wind_turbine}. For the actual data distribution of wind turbine temperature, as shown in Figure \ref{fig:sourcedata}, we use one turbine time series (from five temperature sensors) from January to December in 2017, recorded every 10 minutes, which has around 50,000 observations in total. In order to check concept drifts and data stationary for each variable in actual time-series, we perform the augmented Dickey-Fuller test~\cite{mushtaq2011augmented}, which is used to determine how strongly a series is defined by a trend by calculating the corresponding $p$-value~\cite{dahiru2008p}. The null hypothesis of the test is that the tested series have a certain time-dependent structure (namely not stationary). The $p$-values of the five variables, namely \texttt{Db1t\_avg}, \texttt{Db2t\_avg}, \texttt{Gb1t\_avg}, \texttt{Gb2t\_avg} and \texttt{Ot\_avg}, turn out to be $1.82\times e^{-22}$, $3.34\times e^{-17}$, $3.44\times e^{-20}$, $2.38\times e^{-17}$ and $4\times e^{-6}$, respectively. Since these values are less than $0.05$, we can reject our null hypothesis and conclude that the time series is stationary without concept drifts. We use this actual dataset in our no drift scenario.

\neck{Two synthetic datasets.} In order to evaluate the adaptiveness of our hybrid learning dealing with streaming data concept drifts, we synthetically generate two datasets and simulate gradual drifts and abrupt drifts on each of them, as shown in Figures \ref{fig:gradualdriftdata} and \ref{fig:abruptdriftdata}. 

Let $GD_i(t)$ and $AD_i(t)$ be the generated gradual and abrupt drift value of target variable at timestamp $t$, and $Y_i(t)$ be the true value of input feature, where $i\in [0\ldots n]$. For gradual drift scenario and abrupt drift scenario, the simulation rule for all $n$ variables is specified as Equations \ref{equ:gradual} and \ref{equ:abrupt} separately, where $\alpha_i$ is the \textit{drift value} for variable $i$, $\varepsilon$ is an invariant noise and $\lambda$ is the random \textit{abrupt parameter}.

\begin{equation} \label{equ:gradual}
GD_i(t) = \alpha_i t + Y_i(t) + \varepsilon
\end{equation}
\begin{equation} \label{equ:abrupt}
AD_i(t) = \alpha_i t \lambda + Y_i(t) + \varepsilon
\end{equation}

\subsubsection{Machine learning setting} 
For the evaluation of no-drift, gradual drift and abrupt drift scenario, we split the modeling dataset into training and testing subsets with the ratio of $4:6$. We use 20,000 observations to produce a pre-trained model for batch inference, and send 30,000 observations as streaming data in each time window to test our hybrid learning analytics. The data from all five variables are been normalized using Min-Max Scaling to the range of $[0,1]$ during computation. 

\begin{figure}[h]
\begin{center}
    \includegraphics[width=0.48\textwidth]{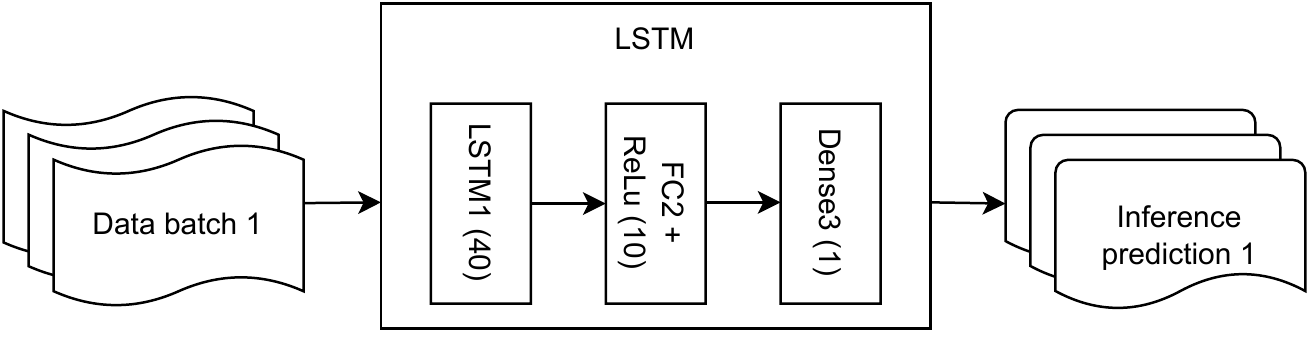}
    \caption{LSTM network architecture.}
    \label{fig:lstm}
\end{center}
\end{figure}

\neck{Settings for model training.} We run a multilayer-perception long short-term memory (LSTM) network shown in Figure \ref{fig:lstm}, which has one long short-term memory layer with 40 units, one fully connected layer with 10 units and ReLu activation, and one final output layer (10,981 total parameters). For the pre-train model used in batch inference, we train the model using 50 epochs and 512 batch sizes with a 0.001 learning rate. For the speed model in each time window, we use 100 epochs and 64 batch sizes with a 0.001 learning rate. Because this study focuses on hybrid learning and its deployment on edge-cloud resources, we did not employ more complicated RNN deep learning models, which can be easily evaluated in future work. 

\neck{Settings for model inference.} Our batch inference loads the pre-trained model in each time window and makes predictions for the records in the current time window. We set the time window size equal to 30 seconds in all experiments and throttle no less than 200 records in each time window. For speed inference, we have three parallel processes, which include 1) fetching the latest pre-trained model from S3 and saving it to the edge disk; 2) loading the latest pre-trained model from the edge disk and making a prediction for the current time window; and 3) training a new LSTM model based on the current time window and uploading the model to S3. Because the three processes run in parallel, we cannot guarantee to use the model trained from the previous time window to infer the current time window. But this approach can improve the latency greatly. For the hybrid inference, the predicted value of each record is calculated from its batch and speed inference prediction.

As the problem statement in Section \ref{sec:statement}, assuming values for $y^{i-1}_{t+1}$, $y^{i-2}_{t+1}$, ..., $y^{i-n}_{t+1}$ are known when making predictions (same with batch inference without the time window $t$). We evaluate the prediction performance by calculating RMSE between predictions $\hat{y}^{i}_{t+1}$ and actual observations $y^{i}_{t+1}$. In the paper, we set the time lag to be 5, namely $n=5$.

\begin{table*}[ht]
\centering
\footnotesize
\caption{The latency of inference phase for different stream analytics with three deployment modalities (unit: second).}
\setlength{\tabcolsep}{0.98mm}{
\begin{tabular}{|c|c|c|c|c|c|c|c|c|c|}
\hline
\multirow{2}{*}{}& \multicolumn{3}{c|}{\textbf{Speed Inference}} & \multicolumn{3}{c|}{\textbf{Batch Inference}} & \multicolumn{3}{c|}{\textbf{Serving/Hybrid Inference}}\\ \cline{2-10}
& ~Computation~ & ~Communication~ & ~Total~ & ~Computation~ & ~Communication~ & ~Total~ & ~Computation~ & ~Communication~ & ~Total~ \\
\hline
Cloud Centric & 8.82 & 13.88 & 22.70 & 8.49 & 14.52 & 23.01 & 23.65 & 16.47 & 40.12 \\ 
\hline
Edge Centric & 10.25 & 6.83 & 17.08 & 10.65 & 6.61 & \textbf{17.26} & 27.19 & 8.52 & 35.71 \\ 
\hline
~Edge-Cloud Integrated~ & 9.89 & 6.75 & \textbf{16.64} & 10.83 & 7.20 & 18.03 & 26.71 & 8.93 & \textbf{35.64} \\ 
\hline
\end{tabular}}
\label{tab:latency_inference}
\end{table*}

\subsubsection{Hardware and software setting} 
\neck{\\Hardware settings.} For our experiments, we use a Raspberry Pi 4 as our front-end on-premise edge device, which is attached with the 32GB MicroSD memory card and 4GB RAM. We use Amazon Web Services as our back-end cloud platform. A data analytics server is deployed on AWS EC2, which allocate to a compute-optimized c5.4xlarge instance with 16 virtual CPUs (vCPUs) and 32GB of memory.

\neck{Software settings.} For the software environment on the edge, we use Debian 11 Bullseye OS with Python 3.8. The dependencies on edge include Tensorflow-lite 2.5, Spark 3.0 and Pandas for inference learning, Kafka 3.1 for data injection, and AWS SDK Boto3 for edge-cloud data and model synchronization. The Kafka data injection bandwidth is around 7 records/second in our experiments. Meanwhile, the software environment on the cloud is encapsulated in our public Docker image, which contains Tensorflow 2.2, Spark 3.0 and Pandas for model training and also Boto3 for synchronization. Both our software environments support the Spark big data analytics engine, which enables parallel computation on two sides.

\subsection{Latency Evaluation for Different Deployment Modalities} \label{sec:6.2}
We first evaluate the performance of hybrid learning framework with the three deployment types explained in Section \ref{sec:flexible}. Since the deployment modalities, including edge-centric, cloud-centric and edge-cloud integrated, only change the resources where the modules are deployed in, the stream analytics still executes based on the same logic which results in the same accuracy performance. Therefore, we only evaluate their latency. 

\begin{figure*}[ht]
	\centering
	\subfloat[Static weighting.]{
		\includegraphics[width=0.35\textwidth]{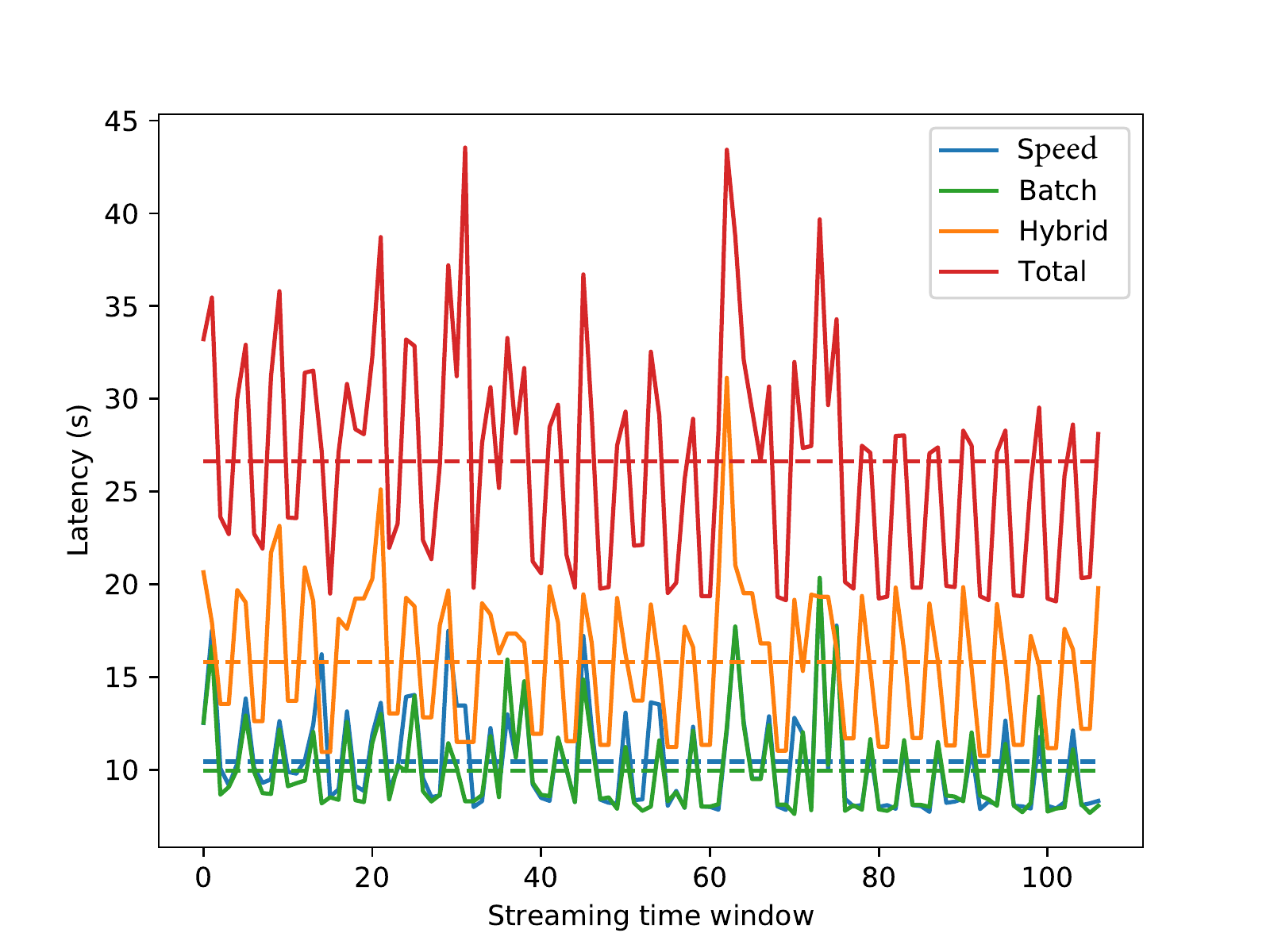}
		\label{fig:fixweights_latency}
	}
	\subfloat[Dynamic weighting.]{
		\includegraphics[width=0.35\textwidth]{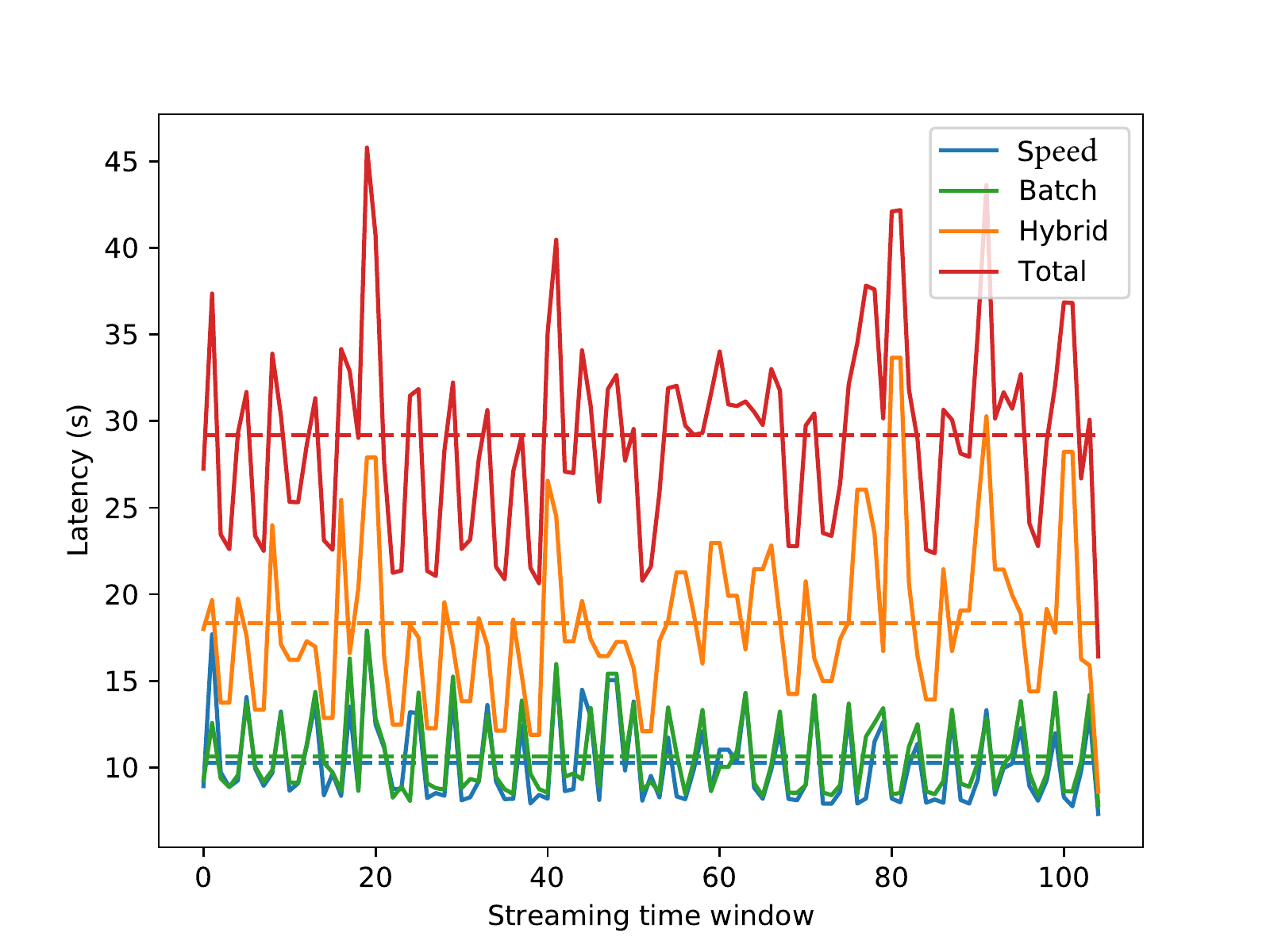}
		\label{fig:bestweights_latency}
	}
	\caption{Inference and total latency of hybrid stream analytics for edge-cloud integrated deployment.}
	\label{fig:latency}
\end{figure*}

We separate the pipeline of stream analytics into two phases: inference phase and training phase. These two phases work asynchronously as illustrated in Figure \ref{fig:workflow}. In inference phase, starting from data injection to prediction archiving, we record both computation latency and communication latency for each time window, then we calculate their averages over all time windows and show the results in Table \ref{tab:latency_inference}. The table shows edge-centric and edge-cloud integrated deployment are more efficient than cloud-centric deployment because of their small communication overheads. For edge-centric and edge-cloud integrated deployment, referred to as Figures \ref{fig:flexible}a and \ref{fig:flexible}b, they have roughly the same latency in inference phase since their module deployments are exactly the same, except the speed training. Next in the training phase, starting from data injection to model synchronization, we only measure the average computation and the communication latency for speed layer, since batch layer only trains a model once and hybrid layer does not have the training phase. For cloud-centric deployment, the average latency of speed layer are 14.73 seconds for computation, 14.47 seconds for communication, and 29.20 seconds in total. For the edge-cloud integrated deployment, the average latency are 15.69, 14.04 and 29.73 seconds, respectively. These two deployment modalities perform in the same trend since their speed training modules are both deployed in cloud. For edge-centric deployment, referred to as Figure \ref{fig:flexible}a, the speed training module should be deployed in edge resource. We also evaluated the edge-centric deployment with our Raspberry Pi edge device, but the experiment failed with out-of-memory error. It shows the edge device with a limited capacity cannot support this type of deployment. So, if we compare the total latency (inference and training), edge-cloud integrated deployment is the best.

In summary, for three deployment modalities, edge-cloud integrated deployment works best, as its efficiency in inference phase and the sufficient capacity in training phase. Specially, comparing with the other two deployments, the edge-cloud deployment can achieve similar latency performance as edge-centric deployment without worrying about capacity limitations. Therefore, for the rest of our evaluation, we only conduct experiments with the edge-cloud integrated deployment.

\subsection{Latency and Accuracy Evaluation for Different Stream Analytics Approaches}
We focus on both latency and accuracy aspects when evaluating our adaptive hybrid stream analytics approaches. For latency, we measure overhead created by stream analytics. For accuracy, we compare its performance with the proposed dynamic weighting algorithm in different streaming concept drift scenarios. For the dynamic weighting algorithm, we evaluate the stacking of two pre-trained models (one latest speed model and one batch model) as explained in Section \ref{sec:dynamicweights}.

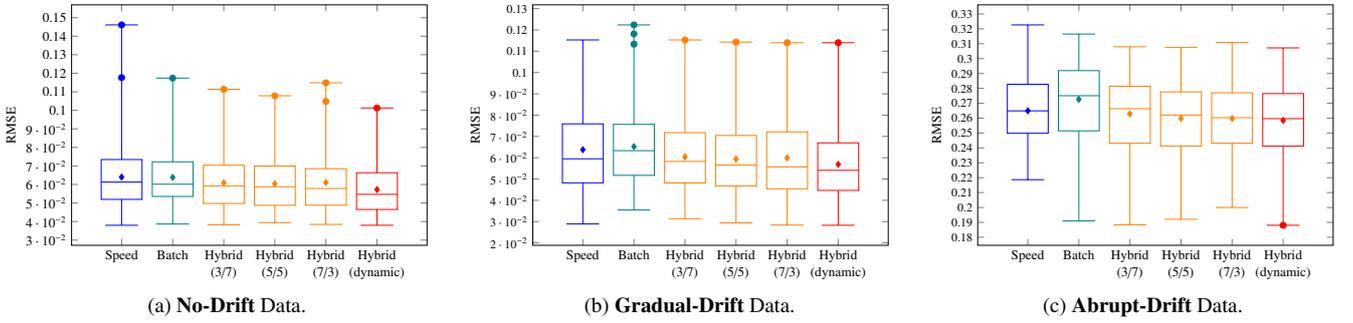
\begin{figure*}[ht]
	\centering
	\subfloat[\textbf{No-Drift} Data.]{
		\input{figures/nodrift_boxplot}
		\label{fig:box-nodrift_fix}
	}
	\subfloat[\textbf{Gradual-Drift} Data.]{
		\input{figures/gradualdrift_boxplot}
		\label{fig:gradualdrift_fix}
	}
	\subfloat[\textbf{Abrupt-Drift} Data.]{
		\input{figures/abruptdrift_boxplot}
		\label{fig:abruptdrift_fix}
	}
	\caption{RMSE box-plots for different inference approaches.}
	\label{fig:rmse_boxplot}
\end{figure*}

\begin{figure*}[ht]
	\centering
	\subfloat[\textbf{No-Drift} Data.]{
		\includegraphics[width=0.33\linewidth]{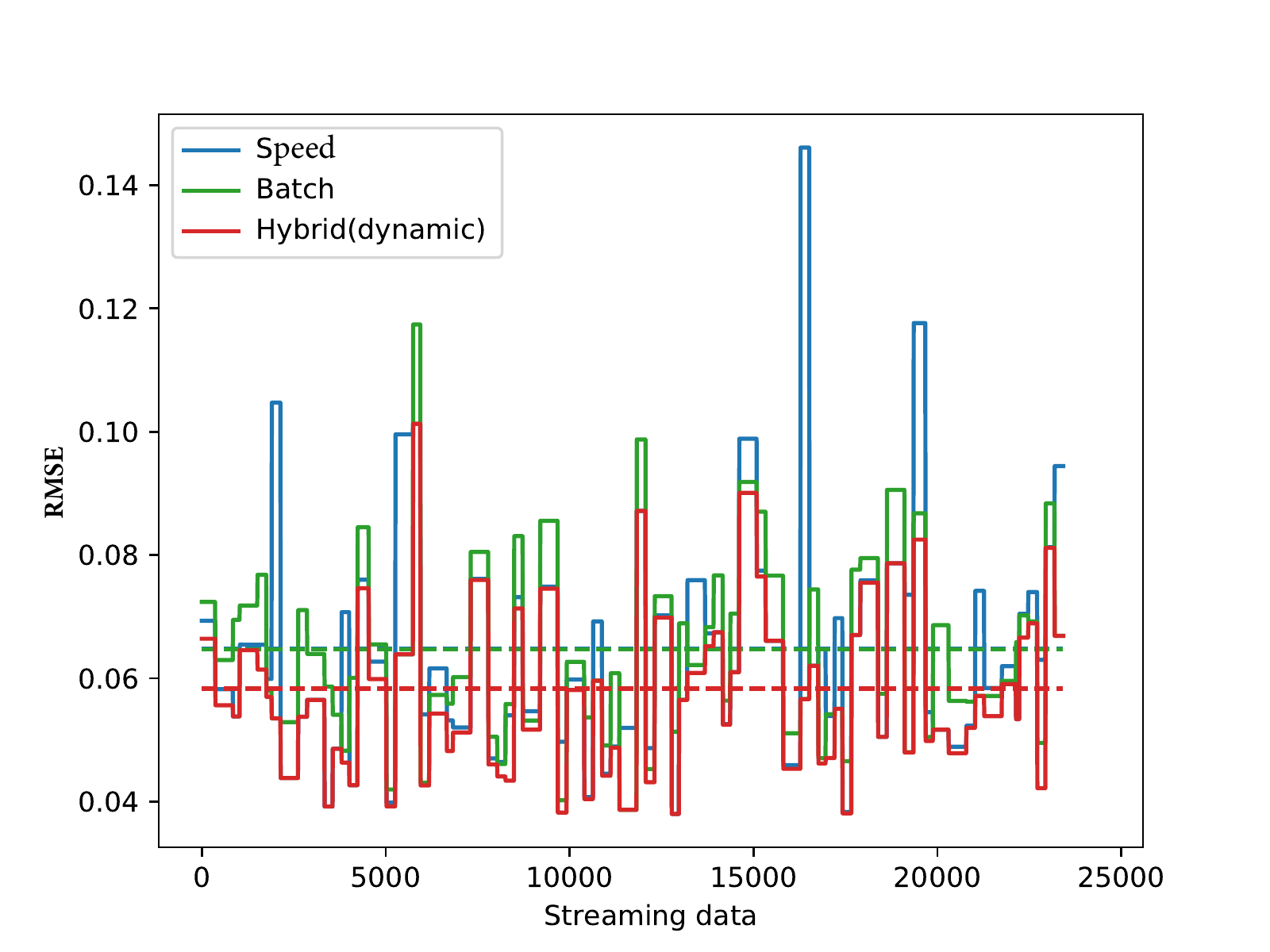}
		\label{fig:nodrift_rmse}
	}
	\subfloat[\textbf{Gradual-Drift} Data.]{
		\includegraphics[width=0.33\linewidth]{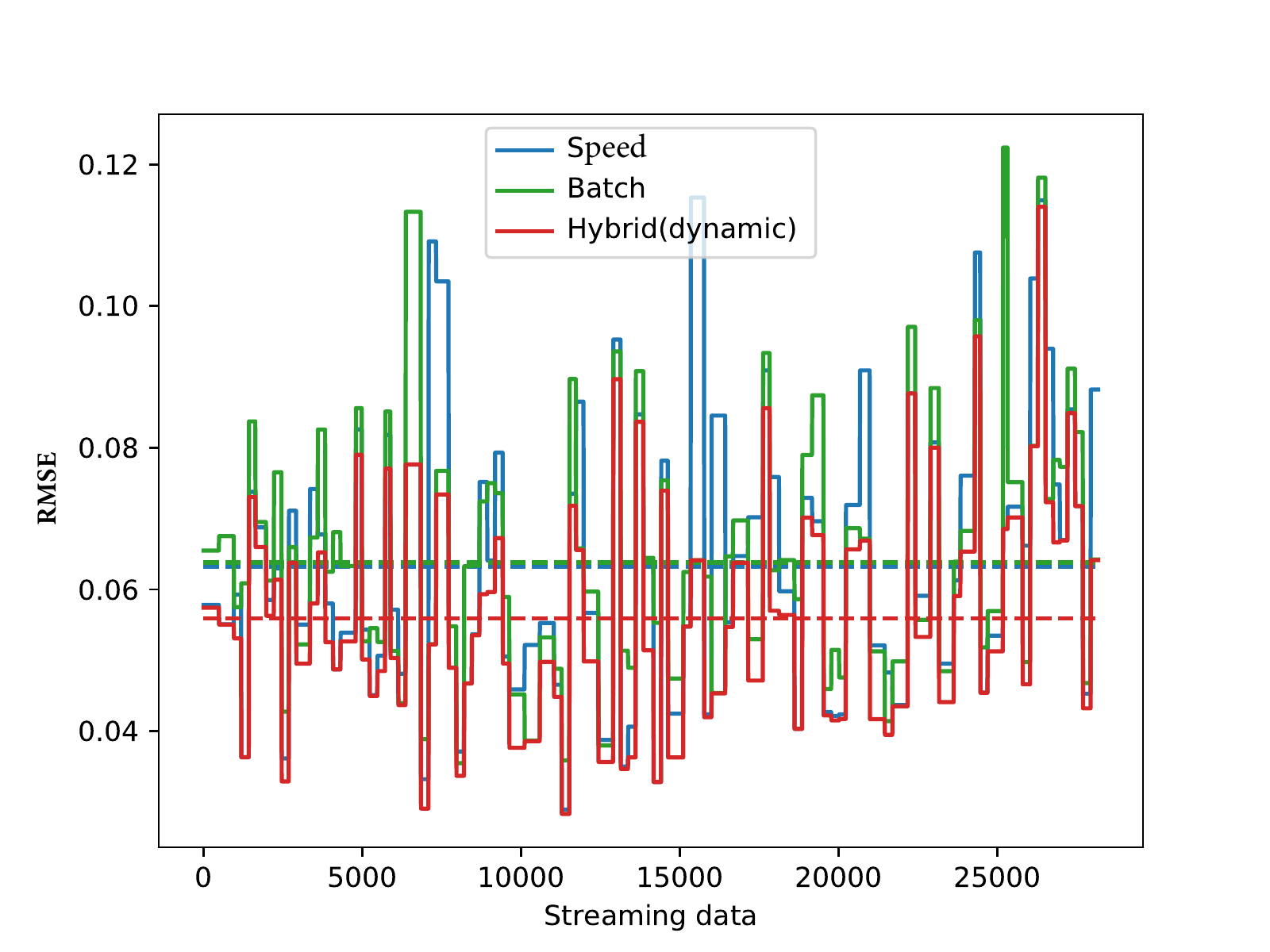}
		\label{fig:gradualdrift_rmse}
	}
	\subfloat[\textbf{Abrupt-Drift} Data.]{
		\includegraphics[width=0.33\linewidth]{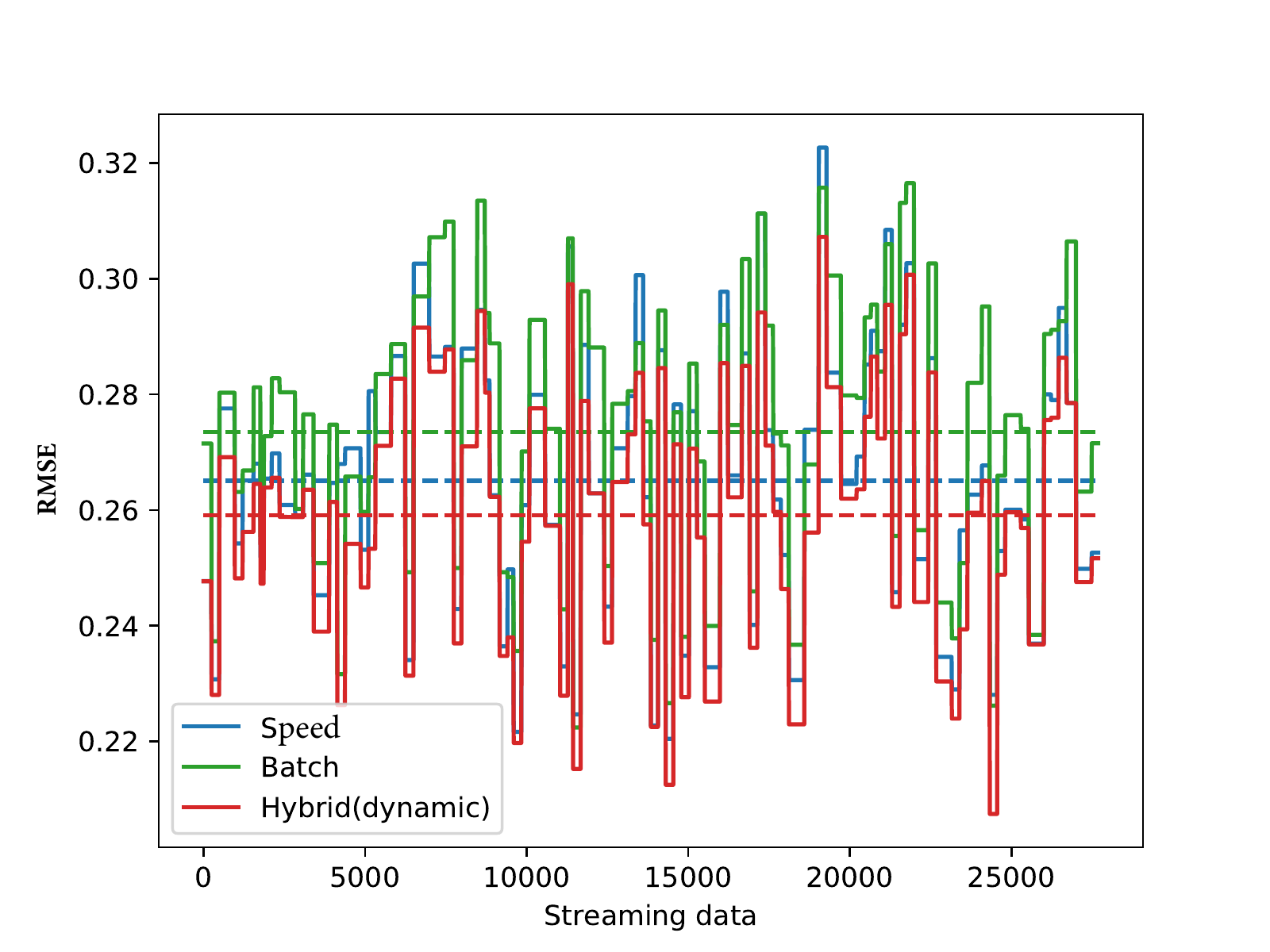}
		\label{fig:abruptdrift_rmse}
	}
	\caption{RMSE of each time window for batch, speed and dynamic hybrid inference.}
	\label{fig:dynamic_rmse}
\end{figure*}

\subsubsection{Latency evaluation}
We first discuss the latency of hybrid stream analytics with static weighting and dynamic weighting. As shown in Figure \ref{fig:latency}, we record the latency of execution in every streaming time window which includes around 200 streaming observations. The latencies of speed and batch inference are in the same trend and they are both lower than the latency of hybrid inference. Since speed and batch inference are executing in parallel and their latencies have overlap, we also evaluate the total latency for the whole hybrid stream analytics. 

With static weighting in Figure \ref{fig:fixweights_latency}, the latencies averaged from all time windows are: 10.43 seconds for speed inference, 9.93 for batch, 15.81 for hybrid and 26.63 for total, respectively. And with dynamic weighting in Figure \ref{fig:bestweights_latency}, the average latencies are 10.25, 10.63, 18.34 and 29.19 seconds, respectively. Since the weight combination algorithm is only applied in hybrid inference, the latencies of speed and batch inference are roughly the same in the two evaluations. For the hybrid inference and the overall hybrid stream analytics, the percentage of average latency increment of dynamic weighting turn out to be 14.82\% and 9.54\%, compared with static weighting. The increment is because our dynamic approach requires time to find the best weights.

\subsubsection{Accuracy evaluation} \label{sec:6.3.2}
To evaluate accuracy performance of hybrid stream analytics, we use Root Mean Squared Error (RMSE) metric to measure how far the predicted values $\hat{y}_j$ are from the ground-truth values $y_j$, as mentioned in Equation \ref{equ:rmse}. We compare the performance of hybrid stream analytics in three data drifting scenarios with two weight combination algorithms. We record the RMSE of inference results for each streaming time window (100 time windows in total), and convert them into the boxplots, as shown in Figure \ref{fig:rmse_boxplot}. For the static weighting algorithm, we also measure the different weights in 3:7, 5:5 and 7:3 (speed:batch) of the accuracy performance evaluation.

Figure \ref{fig:rmse_boxplot} assesses the accuracy of hybrid stream analytics and baseline approaches. In summary, for hybrid stream analytics, both static and dynamic weighting algorithm achieve better RMSE values than speed and batch inference. For no-drift scenario, the batch and speed inference get roughly the same RMSE since there are no unexpected changes in the concept or distribution or the streaming data. For both gradual-drift and abrupt-drift scenarios, the speed inference works better than the batch inference since the latter cannot detect the changes in data distribution based on the model from historical data. On the contrary, speed inference updates the model for each time window, using a model trained from the previous time window to test the streaming data in the current time window, which can catch the drifting in time. Besides, this evaluation also shows the hybrid stream analytics with dynamic weighting algorithm achieves the best average RMSE in all three scenarios, and the improved percentages of RMSE are 10.73\%, 12.73\% and 5.20\% respectively. We also record the RMSE for each time window in all three drifting scenarios, as shown in Figure \ref{fig:dynamic_rmse}. It also shows our dynamic hybrid approach achieves the best RMSE for most time windows.

\begin{table}[ht]
\centering
\footnotesize
\caption{Time percentage of each inference being the best in terms of RMSE for no-drift data.}
\setlength{\tabcolsep}{0.99mm}{
\begin{tabular}{|c|c|c|c|c|}
\hline
\textbf{}  & \textbf{~Static (3:7)~} & \textbf{~Static (5:5)~} &\textbf{~Static (7:3)~}& \textbf{~Dynamic~}\\
\hline
\hline
~~Speed~~ & \textbf{0.5460} & \textbf{0.4757} & 0.3311 & 0.1648 \\
\hline
~Batch~ & 0.1172 & 0.1967 & 0.2578 & 0.088 \\
\hline
~Hybrid~ & 0.3368 & 0.3275 & \textbf{0.4111} & \textbf{0.7472} \\
\hline
\end{tabular}}
\label{tab:nodrift_rmse_percentage}
\end{table}

\begin{table}[ht]
\centering
\footnotesize
\caption{Time percentage of each inference being the best in terms of RMSE for gradual-drift data.}
\setlength{\tabcolsep}{0.99mm}{
\begin{tabular}{|c|c|c|c|c|}
\hline
\textbf{}  & \textbf{~Static (3:7)~} & \textbf{~Static (5:5)~} &\textbf{~Static (7:3)~}& \textbf{~Dynamic~}\\
\hline
\hline
~~Speed~~ & 0.3472 & 0.2830 & 0.1702 & 0.0513 \\
\hline
~Batch~ & 0.1093 & 0.1332 & 0.2230 & 0.0252 \\
\hline
~Hybrid~ & \textbf{0.5436} & \textbf{0.5838} & \textbf{0.6068} & \textbf{0.9235} \\
\hline
\end{tabular}}
\label{tab:gradual_rmse_percentage}
\end{table}

\begin{table}[ht]
\centering
\footnotesize
\caption{Time percentage of each inference being the best in terms of RMSE for abrupt-drift data.}
\setlength{\tabcolsep}{0.99mm}{
\begin{tabular}{|c|c|c|c|c|}
\hline
\textbf{}  & \textbf{~Static (3:7)~} & \textbf{~Static (5:5)~} &\textbf{~Static (7:3)~}& \textbf{~Dynamic~}\\
\hline
\hline
~~Speed~~ & 0.4839 & 0.2106 & 0.1129 & 0.0290 \\
\hline
~Batch~ & 0.0000 & 0.0091 & 0.0183 & 0.0000 \\
\hline
~Hybrid~ & \textbf{0.5161} & \textbf{0.7802} & \textbf{0.8687} & \textbf{0.971} \\
\hline
\end{tabular}}
\label{tab:abrupt_rmse_percentage}
\end{table}

\begin{table*}[ht]
\centering
\footnotesize
\caption{Related works that support different of capabilities. \Circle No \CIRCLE Yes \LEFTcircle Some approaches support it.}
\setlength{\tabcolsep}{0.99mm}{
\begin{tabular}{|c|c|c|c|c|c|}
\hline
\multirow{2}{*}{Approaches}  & \textbf{Inference} & \textbf{Periodic model} & \textbf{Model update}& \textbf{Multi-model analytics}& \textbf{Flexible edge-} \\
& \textbf{on edge} & \textbf{training on cloud} & \textbf{from cloud to edge}& \textbf{for adaptability}& \textbf{cloud deployment} \\
\hline
\hline
Machine learning based IoT stream & \multirow{2}{*}{\LEFTcircle} & \multirow{2}{*}{\Circle} & \multirow{2}{*}{\Circle} & \multirow{2}{*}{\CIRCLE} & \multirow{2}{*}{\Circle} \\
data analytics~\cite{pandya2020adaptive,shao2021adaptive,puschmann2016adaptive,yang2021lightweight,puschmann2016adaptive,yang2021pwpae,mehmood2021concept} &&&&&\\
\hline
Edge-cloud integrated & \multirow{2}{*}{\LEFTcircle} & \multirow{2}{*}{\CIRCLE} & \multirow{2}{*}{\LEFTcircle} & \multirow{2}{*}{\Circle} & \multirow{2}{*}{\Circle} \\
framework~\cite{balouek2019towards,nezami2021decentralized,naha2020deadline,abdelbaky2017computing,murwantara2021adaptive,luckow2021pilot,osia2018private,9615824}&&&&&\\
\hline
System or toolkit for deep learning & \multirow{2}{*}{\CIRCLE} & \multirow{2}{*}{\Circle} & \multirow{2}{*}{\Circle} & \multirow{2}{*}{\Circle} & \multirow{2}{*}{\CIRCLE} \\
based inference~\cite{deepstream,coral,edgetpu,greengrass,iotedge} &&&&&\\
\hline
\textbf{Edge-cloud} &\multirow{2}{*}{\CIRCLE}&\multirow{2}{*}{\CIRCLE}&\multirow{2}{*}{\CIRCLE}&\multirow{2}{*}{\CIRCLE}&\multirow{2}{*}{\CIRCLE}\\
\textbf{integrated framework}&&&&&\\
\hline
\end{tabular}}
\label{tab:related_work}
\end{table*}

In order to verify above conclusions, we further draw Tables \ref{tab:nodrift_rmse_percentage}, \ref{tab:gradual_rmse_percentage} and \ref{tab:abrupt_rmse_percentage}, which show the percentage of each inference being the best with no-drift, gradual-drift and abrupt-drift data separately. For no-drift scenario, hybrid inference is the best approach only in dynamic weights and one static weighting (7:3) algorithm. For both the gradual-drift and abrupt-drift scenarios, however, hybrid inference is always the best approach for every weight combination algorithm. As a result, our hybrid stream analytics can adapt the gradual and abrupt concept drift effectively.

\section{Related work} \label{sec:relatedworks}
There have been many studies related to the topics we discussed in the paper. Based on their framework capabilities and main focused challenges, as shown in Table \ref{tab:related_work}, we categorize them into different groups. Next, we describe the details of these related work from the three aspects we listed in the table.

\subsection{Machine Learning based IoT Stream Data Analytics}
When dealing with concept drift in stream data analytics, we study how to optimally integrate batch learning and stream learning in our paper. Paper~\cite{pandya2020adaptive} proposes a framework that generates improved time series forecasting by supporting batch-based, stream-based and hybrid time series forecasting, to tackle the adaptability challenges. Several papers~\cite{shao2021adaptive,puschmann2016adaptive,yang2021lightweight,rocher2021iohmm,uchiteleva2021trils,yang2021pwpae,mehmood2021concept,zhang2020flexible} study how to update the model based on streaming data and propose their solutions. Shao et al.~\cite{shao2021adaptive} propose an adaptive strategy in conjunction with ensemble learning for the task of concept drift detection, while Puschmann et al.~\cite{puschmann2016adaptive} use an online clustering mechanism to cluster the streaming data, which remains adaptive to drifts by adjusting itself as the data changes. Yang et al.~\cite{yang2021lightweight} propose their drift adaptation method algorithm based on the combination of sliding and adaptive window-based methods, as well as performance-based methods. All these studies focus on the algorithm design for training and updating one identical ML/DL model to deal with streaming concept drift with the best performance. Some of these~\cite{puschmann2016adaptive,yang2021pwpae,mehmood2021concept} did not consider the computational power of the edge computing environment, so the algorithms need to be further deployed in additional resources (like storage or computation optimized device). In contrast, the adaptive hybrid stream analytics we proposed is a weight combination solution from two (batch and stream) inferences, which does not re-evaluate or adjust the layers of the neural network based on the results. By periodically applying knowledge from two trained and compressed models, our work is more lightweight and portable for edge devices in real-time concept drift adaptation.

\subsection{Edge-cloud Integration}
There are also several studies~\cite{balouek2019towards,nezami2021decentralized,naha2020deadline,abdelbaky2017computing,murwantara2021adaptive} for workload management on edge-cloud integrated resources, more targeting an optimized cyber-infrastructure design but not much for machine learning related workload. Luckow et al.~\cite{luckow2021pilot} study how to manage distributed edge and cloud resources and studies the performance of machine learning models for the outlier detection. However, the edge devices in the paper are simulated so it is yet to be seen whether the findings are the same in real-world. Also, the machine learning models are only deployed on the cloud side, not on the edge side. Osia et al.~\cite{osia2018private} deploy deep learning models to predict images collected by the edge, where sensitive information is first pre-processed on the edge and its representation is sent to the cloud for complex inferences. Since the edge devices are only used for data pre-processing, not for actual machine learning based inference, the total latency of their framework will be higher than inferring directly on edge. Abdulla et al.~\cite{9615824} argue that using adaptive learning for streaming data processing could solve concept drift problems, and the proposed cooperative fog-cloud architecture shows updating machine learning models periodically can help reduce RMSE by about 20\%. However, their experiments did not use portable edge devices such as Raspberry Pi, instead they used a local computer. The edge-cloud integrated framework we proposed is a general and flexible design, whose docker-based modules can be developed in either cloud or edge side even with different types of edge devices.

\subsection{Complete System or Toolkit for Deep Learning based Inference}
There have been many systems or toolkits that support deep learning based inference on IoT/edge devices. NVIDIA's DeepStream~\cite{deepstream} is a streaming analytic toolkit that helps the user build and deploy video analytics applications on-premises, on the edge, and in the cloud. DeepStream features hardware-accelerated building blocks~\cite{deepstreamsdk} that bring deep neural networks and other complex stream data processing tasks into GStreamer processing pipelines and maximize the computation using GPUs. Based on its design, DeepStream can be highly optimized to run on NVIDIA series or GPU-enabled edge devices like Jetson Xavier NX and Jetson Nano. However, while DeepStream has multiple examples that are provided as source code, its SDK is not released as open-source software. An alternative toolkit for deep learning based inference is Google Coral~\cite{coral}. Coral is a complete toolkit for building intelligent devices with fast deep neural network inferencing. Same with DeepStream, Coral can enable its peak capability with the proposed hardware and software solutions like Edge TPU coprocessor~\cite{edgetpu}. 

More focused on deployment rather than inference learning, some Cloud platforms like AWS and Azure also provide their general-propose solution for edge devices inference even offline from the cloud. AWS IoT Greengrass~\cite{greengrass} is an open-source edge runtime and cloud service for building, deploying, and managing edge devices. Greengrass manages and operates multiple edge devices in the field locally or remotely using MQTT or other protocols. With the solution, inference can be deployed across edges using any language, packaging technology, or runtime. Our hybrid learning framework is based on the Greengrass runtime. We further study how to deploy stream analytics among edge and cloud resources and improve their accuracies. Microsoft also provides Azure IoT Edge~\cite{iotedge} service to scale out inference learning by packaging the logic into standard containers, deploying these containers to any of the edge devices and monitoring it all from the cloud. Different from Greengrass, the applications like inference learning in Azure IoT Edge need to be developed in one of the supported programming languages.

\section{Conclusions} \label{sec:conclusions}
Stream analytics aims to analyze and process high volumes of streaming data continuously. In this paper, we study how to best leverage edge and cloud resources to achieve better accuracy and latency for RNN-based stream analytics and better adapt concept drift in stream data. We propose three flexible deployments for the hybrid stream analytics framework in order to achieve the proper trade-off between latency and accuracy for stream analytics. We also propose an adaptive and dynamic hybrid learning model with two weight combination algorithms for solving the concept drift during stream analytics. The evaluation with real-world stream datasets shows the proposed edge-cloud deployment can archive similar latency performance as edge-centric deployment without worrying about capacity limitations, and our dynamic weighting algorithm performs the best among other hybrid learning model approaches for all three concept drift scenarios in terms of accuracy.

For future work, we will mainly focus on the following three aspects of the hybrid stream analytics framework. First, the Spark-based speed training module can be extended to multiple edge devices as a distributed master-worker computing. Second, we will study more variants of the proposed dynamic weighting algorithm, like stacking the most resent $n$ speed layer models or stacking speed layer models continuously. Last, we will try more advanced RNN deep learning models like attention models with our framework.

\section*{Acknowledgment}
This work is supported by the National Science Foundation (NSF) Grant No. OAC--1942714 and U.S. Army Grant No. W911NF2120076.

\bibliographystyle{elsarticle-num}
\bibliography{ref}

\end{document}

%% file: figures/nodrift_boxplot.tex
\begin{tikzpicture}[scale=0.56]
  \begin{axis}
    [
    ylabel={RMSE},
    boxplot/draw direction=y,
    ytick distance=0.01,
    xtick={1,2,3,4,5,6},
    xticklabels={Speed, Batch, Hybrid\\(3/7), Hybrid\\(5/5), Hybrid\\(7/3), Hybrid\\(dynamic)},
    x=1.2cm,
    ticklabel style = {font=\small},
    label style={font=\small},
    cycle list={},
    x tick label style={font=\small, text width=3cm, align=center}
    ]
    \addplot+[color=blue,mark=*,boxplot prepared={
      median=0.061306071279906,
      average=0.064,
      upper quartile=0.073457462484836,
      lower quartile=0.05197341002243,
      upper whisker=0.14610415604688515,
      lower whisker=0.03800404772480368
    },
    ] coordinates {(1,0.14610415604688515)(1,0.1176377711224592)}
    ;
    \addplot+[color=teal,mark=*,boxplot prepared={
      median=0.060143133963061,
      average=0.06383443526305,
      upper quartile=0.072248712512195,
      lower quartile=0.053546404208206,
      upper whisker=0.1174079608500742,
      lower whisker=0.03867927515794907
    },
    ] coordinates {(2,0.1174079608500742)};
    \addplot+[color=orange,mark=*,boxplot prepared={
      median=0.05923194112625,
      average=0.060870383881735,
      upper quartile=0.070386717664949,
      lower quartile=0.049746985022322,
      upper whisker=0.11136590189080087,
      lower whisker=0.038235008142040665
    },
    ] coordinates {(3,0.11136590189080087)};
    \addplot+[color=orange,mark=*,boxplot prepared={
      median=0.058768793835576,
      average=0.06038610507721,
      upper quartile=0.069961710167574,
      lower quartile=0.048771370080705,
      upper whisker=0.10786738004861031,
      lower whisker=0.039419569642652114
    },
    ] coordinates {(4,0.10786738004861031)};
    \addplot+[color=orange,mark=*,boxplot prepared={
      median=0.057847481843289,
      average=0.061078469457309,
      upper quartile=0.068508789072178,
      lower quartile=0.048820600233801,
      upper whisker=0.11487546179678669,
      lower whisker=0.038456006469200554
    },
    ] coordinates {(5,0.11487546179678669)(5,0.10484270791003526)};
    \addplot+[color=red,mark=*,boxplot prepared={
      median=0.054679115400268,
      average=0.057252620998843,
      upper quartile=0.066342705784507,
      lower quartile=0.046521484787144,
      upper whisker=0.1012852532003858,
      lower whisker=0.03800404772480368
    },
    ] coordinates {(6,0.1012852532003858)};
  \end{axis}
\end{tikzpicture}

%% file: figures/gradualdrift_boxplot.tex
\begin{tikzpicture}[scale=0.56]
  \begin{axis}
    [
    ylabel={RMSE},
    boxplot/draw direction=y,
    ytick distance=0.01,
    xtick={1,2,3,4,5,6},
    xticklabels={Speed, Batch, Hybrid\\(3/7), Hybrid\\(5/5), Hybrid\\(7/3), Hybrid\\(dynamic)},
    x=1.2cm,
    ticklabel style = {font=\footnotesize},
    label style={font=\small},
    cycle list={},
    x tick label style={font=\small, text width=3cm, align=center}
    ]
    \addplot+[color=blue,mark=*,boxplot prepared={
      median=0.05949924799156,
      average=0.063782921945033,
      upper quartile=0.075904749331151,
      lower quartile=0.048208121196952,
      upper whisker=0.11530735185479453,
      lower whisker=0.02891920539206822
    },
    ] coordinates {};
    \addplot+[color=teal,mark=*,boxplot prepared={
      median=0.063319545042583,
      average=0.065202508923639,
      upper quartile=0.075685237189033,
      lower quartile=0.051725379407986,
      upper whisker=0.12237295607598879,
      lower whisker=0.03545038573225923
    },
    ] coordinates {(2,0.12237295607598879)(2,0.1181197185551361)(2,0.11329659745931635)};
    \addplot+[color=orange,mark=*,boxplot prepared={
      median=0.058259785800284,
      average=0.060456533295734,
      upper quartile=0.07179153923464,
      lower quartile=0.04815245545392,
      upper whisker=0.11530817478601547,
      lower whisker=0.0313987002574371
    },
    ] coordinates {(3,0.11530817478601547)};
    \addplot+[color=orange,mark=*,boxplot prepared={
      median=0.056534654377976,
      average=0.059328197221116,
      upper quartile=0.070459128934184,
      lower quartile=0.046764691585476,
      upper whisker=0.11429816707960112,
      lower whisker=0.029405106944710643
    },
    ] coordinates {(4,0.11429816707960112)};
    \addplot+[color=orange,mark=*,boxplot prepared={
      median=0.055714635356924,
      average=0.059911885677496,
      upper quartile=0.072137923996386,
      lower quartile=0.045426879251016,
      upper whisker= 0.11400292223191644,
      lower whisker=0.028394225299359664
    },
    ] coordinates {(5,0.11400292223191644)};
    \addplot+[color=red,mark=*,boxplot prepared={
      median=0.054098419214037,
      average=0.05690464070773,
      upper quartile=0.06698760330255,
      lower quartile=0.044639474970622,
      upper whisker=0.11399998719496038,
      lower whisker=0.02829156397207287
    },
    ] coordinates {(6,0.11399998719496038)};
  \end{axis}
\end{tikzpicture}

%% file: figures/abruptdrift_boxplot.tex
\begin{tikzpicture}[scale=0.56]
  \begin{axis}
    [
    ylabel={RMSE},
    boxplot/draw direction=y,
    ytick distance=0.01,
    xtick={1,2,3,4,5,6},
    xticklabels={Speed, Batch, Hybrid\\(3/7), Hybrid\\(5/5), Hybrid\\(7/3), Hybrid\\(dynamic)},
    x=1.2cm,
    ticklabel style = {font=\footnotesize},
    label style={font=\small},
    cycle list={},
    x tick label style={font=\small, text width=3cm, align=center}
    ]
    \addplot+[color=blue,mark=*,boxplot prepared={
      median=0.26489993245325,
      average=0.26493231581204,
      upper quartile=0.2827491063119,
      lower quartile=0.24983473433749,
      upper whisker=0.3226389585919533,
      lower whisker=0.2186703916058862
    },
    ] coordinates {};
    \addplot+[color=teal,mark=*,boxplot prepared={
      median=0.27506921370303,
      average=0.27265491926365,
      upper quartile=0.29191784207765,
      lower quartile=0.25147950818855,
      upper whisker=0.31649296394116566,
      lower whisker=0.19100796037932188
    },
    ] coordinates {};
    \addplot+[color=orange,mark=*,boxplot prepared={
      median=0.26636308559419,
      average=0.26288178311104,
      upper quartile=0.28142301784519,
      lower quartile=0.24315606848459,
      upper whisker=0.30794265839908896,
      lower whisker=0.18832989024993457
    },
    ] coordinates {};
    \addplot+[color=orange,mark=*,boxplot prepared={
      median=0.26210164094562,
      average=0.25985093834178,
      upper quartile=0.27764337391807,
      lower quartile=0.24130351268386,
      upper whisker=0.3074736941773727,
      lower whisker=0.19207707072151864
    },
    ] coordinates {};
    \addplot+[color=orange,mark=*,boxplot prepared={
      median=0.26031873122615,
      average=0.25972798449679,
      upper quartile=0.27703988829485,
      lower quartile=0.24315790764763,
      upper whisker=0.31080632543422154,
      lower whisker=0.20000306811715068
    },
    ] coordinates {};
    \addplot+[color=red,mark=*,boxplot prepared={
      median=0.2596563939172,
      average=0.25850139295013,
      upper quartile=0.2764865542339,
      lower quartile=0.24126468064582,
      upper whisker=0.3072019033972688,
      lower whisker=0.18806022026386704
    },
    ] coordinates {(6,0.18806022026386704)};
  \end{axis}
\end{tikzpicture}